\documentclass[12pt,draft]{article}
 
\newcommand{\be}{\begin{equation}}
\newcommand{\ee}{\end{equation}}
\newcommand{\bea}{\begin{eqnarray}}
\newcommand{\eea}{\end{eqnarray}}
\newcommand{\beaa}{\begin{eqnarray*}}
\newcommand{\eeaa}{\end{eqnarray*}}

\newcommand{\hsp}{\hspace{0.1in}}
\newcommand{\del}{\partial}
\newcommand{\g}{{\cal G}}
\newcommand{\B}{{\cal B}}

\newcommand{\F}{{\cal F}}
\newcommand{\J}{{\cal J}}
\newcommand{\jh}{ \hat{\jmath} }
\newcommand{\K}{{\cal K}}
\newcommand{\p}{{\cal P}}
\newcommand{\h}{{\cal H}}
\newcommand{\Lg}{{\bf g}}

\newcommand{\BB}{{{\rm I} \kern -2pt \rlap {\rm B} \kern +8pt}}
\newcommand{\dxy}{\delta(x{-}y)}
\newcommand{\dpxy}{\delta'(x{-}y)}
\newcommand{\half}{{\textstyle{1\over2}}}
\newcommand{\threehalf}{{\textstyle{3\over2}}}
\newcommand{\quarter}{{\textstyle{1\over4}}}

\newcommand{\de}{\delta}
\newcommand{\eps}{\epsilon}
\newcommand{\tr}{{\rm Tr}}
\newcommand{\str}{{\rm sTr}}
\newcommand{\rank}{{\rm rank}}

\advance\textheight by \topskip
\makeatletter

\@addtoreset{equation}{section}

\def\section{\@startsection {section}{1}{\z@}{-3.5ex plus -1ex minus
 -.2ex}{2.3ex plus .2ex}{\large\bf\centering}}
\def\subsection{\@startsection{subsection}{2}{\z@}{-3.25ex plus%
 -1ex minus -.2ex}{1.5ex plus .2ex}{\bf}}
\def\subsubsection{\@startsection{subsubsection}{3}{\z@}{-3.25ex plus%
 -1ex minus -.2ex}{1.5ex plus .2ex}{\sl}}
\makeatother

\begin{document}

\baselineskip 18pt
\parindent 12pt
\parskip 10pt

\begin{titlepage}
\begin{flushright}
PUPT-1908\\
DAMTP-99-176\\
Imperial/TP/99-00/15\\
hep-th/0001222v2\\
January 2000\\[3mm]
\end{flushright}
\vspace{.4cm}
\begin{center}
{\Large {\bf
Conserved charges and supersymmetry\\ in principal chiral and WZW models}}\\
\vspace{1cm}
{\large J.M. Evans${}^{a,b}$\footnote{e-mail: evans@feynman.princeton.edu, 
J.M.Evans@damtp.cam.ac.uk}, 
M. Hassan${}^b$\footnote{e-mail: M.U.Hassan@damtp.cam.ac.uk},
N.J. MacKay${}^c$\footnote{e-mail: nm15@york.ac.uk}, 
A.J. Mountain${}^d$\footnote{e-mail: A.Mountain@ic.ac.uk}}
\\
\vspace{3mm}
{\em ${}^a$ Joseph Henry Laboratories, 
Princeton University, Princeton NJ 08544, U.S.A.}\\
{\em ${}^b$ DAMTP, University of Cambridge, Silver Street, Cambridge
CB3 9EW, U.K.}\\
{\em ${}^c$ Department of Mathematics, University of York, York YO10
5DD, U.K.}\\
{\em ${}^d$ Blackett Laboratory, Imperial College, Prince Consort Road,
London SW7 2BZ, U.K.}\\
\end{center}

\vspace{1cm}
\begin{abstract}
\noindent
Conserved and commuting charges are investigated in both bosonic and 
supersymmetric classical chiral models, with and without Wess-Zumino terms.
In the bosonic theories, there are conserved currents based on 
symmetric invariant tensors of the underlying algebra, and the construction 
of infinitely many commuting charges, with spins equal to the exponents of 
the algebra modulo its Coxeter number, can be carried out irrespective of
the coefficient of the Wess-Zumino term. 
In the supersymmetric models, a different pattern of conserved quantities 
emerges, based on antisymmetric invariant tensors.
The current algebra is much more complicated than in the bosonic case,
and it is analysed in some detail.
Two families of commuting charges can be constructed, each with finitely 
many members whose spins are exactly the exponents of the algebra 
(with no repetition modulo the Coxeter number). 
The conserved quantities in the bosonic and 
supersymmetric theories are only indirectly related, 
except for the special case of the WZW model and its supersymmetric 
extension.
\end{abstract}

\end{titlepage}

\section{Introduction}

In previous work \cite{EHMM2} we investigated local and non-local 
conserved charges in the bosonic principal chiral model (PCM) 
based on a compact classical Lie group. 
The non-local charges, forming the `Yangian' quantum group, were 
shown to commute with the local charges. The algebra of the local 
charges themselves was found to be quite involved in general.
Nevertheless, we were able to prove the existence of mutually commuting 
sets of local charges with spins equal to the 
exponents of the underlying classical Lie algebra modulo its Coxeter number.
This is precisely the structure familiar from affine Toda
field theory (see {\it e.g.\/} \cite{corri94,dorey91}) 
and to this extent the results 
provide a new perspective on properties \cite{chari95}
of the known S-matrices for PCMs \cite{ogie86}. 

In this paper we extend these earlier results in a number of directions.
We begin by adding a Wess-Zumino (WZ) term to the bosonic PCM.
A special case, at a critical value of the coupling, is the
quantum-conformally-invariant WZW model.
We show in section 2 how the results of 
\cite{EHMM2} can be carried over in the presence of a WZ term
with arbitrary coefficient.
A more challenging problem is to find analogous results 
when fermions are added to the PCM (with or without a WZ term) 
so as to make the model supersymmetric,
and it is to this that we devote the remainder of the paper.\footnote{Some 
preliminary results were described in conference talks \cite{EHMM1}.}
A number of novel features arise in conjunction with 
supersymmetry and a proper comparison with the bosonic theory 
is greatly facilitated by a thorough understanding of the 
special behaviour of the critical WZW and super WZW models, 
thus making contact with the results of section 2.

The supersymmetric PCM is introduced in section 3 using 
both superspace and component fields.
We then construct local and non-local conserved charges 
in the model, and explain the effects of adding a WZ term in these
supersymmetric theories.
In section 4 we investigate in detail the algebra of local currents,
before proving the existence of sets of commuting charges whose spins 
are once again related to the exponents of the relevant Lie algebra. 
Various technical or supplementary results are collected in some
appendices.
 
\section{The bosonic principal chiral and WZW models}

\subsection{The lagrangian, symmetries and currents}

The principal chiral model with a Wess-Zumino term can be defined 
by an action \cite{WZW}
\be\label{pcmlagr}
{ \kappa \over 2 } \int d^2 x \, \tr\left( \partial_\mu g^{-1}
 \partial^\mu g\right) 
+ { \kappa \over 3} \, \lambda \int_B \tr (g^{-1} dg )^3 \ .
\ee
We shall refer to this theory 
as the principal-chiral-Wess-Zumino model or PCWZM.
The field $g(x^\mu)$ is a function on spacetime (with coordinates
$x^\mu$) taking values in some compact Lie group
${\cal G}$. This field must then be smoothly extended 
to a three-dimensional manifold $B$ whose boundary is spacetime 
in order to write the WZ term as above (where we have chosen to use 
differential form notation). 
The coupling constants $\lambda$ and $\kappa$ are dimensionless.
Indeed, $\kappa$ is irrelevant classically, and may be assigned any numerical
value without altering the theory. 
The combination $\kappa \lambda$ appearing in the coefficient of the WZ term 
must be quantised in suitable units for a consistent quantum theory,
but this will not be significant for us. 

We shall consider only simple  
classical groups $\g = SU(N)$, $SO(N)$, $Sp(N)$ 
($N$ even in the last case)
with the field $g(x^\mu)$ a matrix in the defining representation.
The corresponding Lie algebra $\Lg$ then consists
of $N{\times}N$ complex matrices $X$ which obey
\bea 
su(N): &\quad & X^\dag = - X, \quad \tr(X) = 0\nonumber \\
so(N): &\quad & X^* = X, \quad \quad X^T = - X \label{SOUP} \\
sp(N): &\quad & X^\dag = -X, \quad X^T = - JXJ^{-1} \nonumber
\eea
where $J$ is some chosen symplectic structure.
In each case we introduce a basis of anti-hermitian generators $t^a$ for 
$\Lg$ with real structure constants $f^{abc}$ and
normalisations given by 
\be 
[ t^a , t^b ] = f^{abc} t^c \; , \qquad \tr (t^a t^b) = -
\delta^{ab} \; .
\ee 
(Lie algebra indices will always be taken from the beginning of the
alphabet.)
For any $X \in \Lg$ we write 
\be\label{alcpts}
X = t^{a} X^a \; ,  \qquad X^a = - \tr(t^a X) \; .
\ee

The advantage of writing the WZ term as a three-dimensional integral is that 
it makes manifest the continuous symmetry 
\be\label{chsym}
{\cal G}_L \times {\cal G}_R \; : \qquad 
g \mapsto U^{\phantom{1}}_{\rm L} \,g\, U^{-1}_{\rm R} \; .
\ee
It is convenient to introduce the quantities 
\be
E^L_\mu = \del_\mu g g^{-1} \, , \qquad 
E^R_\mu = - g^{-1} \del_\mu g
\ee
which take values in the Lie algebra ${\bf g}$
and transform only under the left and right factors of the symmetry
group respectively. 
The equations of motion from the action (\ref{pcmlagr}) above state 
that the currents 
\be\label{lrcurr} 
j_\mu^L= 
{\kappa} (E^L_\mu - \lambda \varepsilon_{\mu \nu} E^{L \, \nu} ) , \,
\qquad 
j_\mu^R= {\kappa} 
(E^R_\mu + \lambda \varepsilon_{\mu \nu} E^{R \, \nu} ) 
\ee
are conserved:
\be\label{pcwzm}
\del^\mu j_\mu^{R} = 
\del^\mu j_\mu^{L} = 0 \, .
\ee

We shall use both orthonormal coordinates
$x^0 = t$ and $x^1 = x$ in two dimensions, 
with conventions $\eta^{00} = - \eta^{11} = 1$, 
$\varepsilon^{01} = 1$, and also 
light-cone coordinates and derivatives defined by
\[
x^\pm = {1\over 2} (t \pm x) \, , \qquad \del_\pm = \del_t \pm \del_x
\, .
\] 
In terms of the latter, the definitions of the conserved currents become
\be
j^L_\pm = \kappa ( 1 \mp \lambda ) E^L_\pm
\, , \qquad
j^R_\pm = \kappa ( 1 \pm \lambda ) E^R_\pm
\ee
and the equations of motion are equivalent to 
either of the conditions
\be
\label{eqnmot}
\del_\mp j^L_\pm = \mp {\kappa \over 2} [ j^L_+ , j^L_- ] \, , \qquad
\del_\mp j^R_\pm = \mp {\kappa \over 2} [ j^R_+ , j^R_- ]
\, .
\ee

The PCM is of course a special case of the PCWZM corresponding 
to the choice $\lambda = 0 $. 
WZW models are defined by the conditions $\lambda = \pm 1$,
which we shall refer to as critical points, 
or critical values of the coupling (though we shall concentrate 
on the classical theories in this paper). 
The special nature of these critical points 
is evident in light-cone coordinates,
since then
$j^L_\pm = j^R_\mp = 0$ and the equations of motion become simply
\be\label{wzw} 
\del_\mp j^R_\pm = \del_\pm j^L_\mp = 0
\ee
We shall refer to conservation equations of this special type as
{\it holomorphic}.

For our purposes it will be sufficient to deal with just one of the 
currents $j_\mu^L$ or $j_\mu^R$; we choose the right current and drop 
the label $R$ henceforth.

The classical action is conformally-invariant and as a result the
energy-momentum tensor
\be 
T_{\mu \nu} = -{1 \over 2 \kappa} \left ( \, 
\tr (j_\mu j_\nu) - {1\over2}\eta_{\mu \nu} \tr
 (j_\rho j^\rho) \, \right )
\ee
is not only conserved and symmetric but also traceless.
In light-cone components it takes the familiar form 
\be 
T_{\pm \pm} = -{1 \over 2 \kappa } 
\tr ( j_{\pm} j_{\pm}) \; , \qquad T_{+-} = T_{-+} = 0
\; ,
\ee
with 
\be\label{conf}
\partial_- T_{++} = \partial_+ T_{--} = 0 \; .
\ee
The WZ term does not contribute to the energy-momentum tensor because it
is metric-independent.

Finally we remark on the discrete transformation
\be\label{parity}
\pi \, : \; g\mapsto g^{-1} \qquad \Rightarrow
 \qquad E_\mu^L \leftrightarrow E_\mu^R \, , 
\ee
which exchanges $\g_L$ and $\g_R$ and which we shall consequently 
refer to as $\g$-parity.
For the PCM, with $\lambda = 0$, 
this is a symmetry of the lagrangian.
For the more general case of the PCWZM it is not a symmetry by itself,
but it is if combined with the usual spacetime parity transformation 
$x \mapsto -x$. 

\subsection{Poisson brackets}

The canonical Poisson brackets of  
the conserved currents in the PCWZM are 
\begin{eqnarray}
\{j_0^a(x),j_0^b(y)\} & =  & f^{abc}j_0^c(x) \dxy + 
2\lambda \kappa \delta^{ab} \dpxy \nonumber \\
\{j_0^a(x),j_1^b(y)\} & =  & f^{abc}j_1^c(x) \dxy + 
(1{+}\lambda^2) \kappa \delta^{ab} \dpxy \label{STPBs} \\
\{j_1^a(x),j_1^b(y)\} & =  & f^{abc}(\, 2\lambda j_1^c(x) -
\lambda^2 j_0^c(x) \, ) \, \delta(x{-}y) + 2 \lambda \kappa \delta^{ab} \dpxy 
\nonumber
\end{eqnarray}
or, in light-cone coordinates,
\begin{eqnarray}
\{j_\pm^a(x),j_\pm^b(y)\} & =  & \half f^{abc}(1\pm\lambda)\left(  
( 3\mp\lambda) j_\pm^c(x) - (1\pm\lambda) j_\mp^c(x) \right) \delta(x{-}y)
\nonumber \\ 
&&\pm 2 \kappa (1\pm\lambda)^2 \delta^{ab} \delta'(x{-}y) \label{LCPBs} \\
\{j_+^a(x),j_-^b(y)\} & =  & \half f^{abc}\left( 
(1-\lambda)^2 j_+^c(x)  + (1+\lambda)^2 j_-^c(x) \right)
\dxy \nonumber
\end{eqnarray}
We repeat that we are now dealing exclusively with the $R$ currents 
(the brackets of the $L$ currents with themselves are similar, while those
of $L$ with $R$ are more complicated, but we shall need 
neither).
These brackets may be derived in various ways; 
one method is sketched in an appendix, section 6.

Notice that different values of $\lambda$ result in genuinely 
different current algebras. 
In particular, we observe that at the critical values 
$\lambda= \pm 1$, with $j_\mp = 0$, the surviving 
current component $j_\pm$ obeys a Kac-Moody algebra.
The value of $\kappa$, on the other hand, is of no real significance.
Corresponding to its appearance as an overall 
factor in the lagrangian, it could be eliminated from the current algebra
by a simultaneous re-scaling of currents and Poisson brackets.
We can take advantage of this when carrying out calculations,
by setting $\kappa$ to some convenient numerical value.

\subsection{Local conserved charges and invariant tensors}

Consider the PCWZM based on $\g$ with arbitrary couplings $\kappa$ and 
$\lambda$.
Let $d^{(m)}_{a_1 a_2 \ldots a_m}$ be any totally symmetric invariant tensor, 
(we shall not always indicate the rank $m$ explicitly) so that
\be
d_{c(a_1a_2 \ldots a_{m-1}}f_{a_m)bc} = 0 \ .
\label{dinv}\ee
For each such $d$-tensor there are holomorphic 
conservation equations
\be 
\label{gencons}
\del_-( \, d_{a_1 a_2 \ldots a_m} j_+^{a_1} j_+^{a_2} \ldots
j_+^{a_m} \, ) = 
\del_+( \, d_{a_1 a_2 \ldots a_m} j_-^{a_1} j_-^{a_2} \ldots
j_-^{a_m} \, ) = 0 \ , 
\ee
which follow immediately from (\ref{dinv}) and from (\ref{eqnmot}) 
written in the form 
\[ 
\del_\mp j_\pm^a = \mp \half \kappa f^{abc} j_+^b j_-^c \ .
\]

One special case is 
\be\label{confpower}
\partial_- (T^n_{++}) = \partial_+ (T_{--}^n) = 0 
\ee
which corresponds to  
even-rank invariant tensors constructed from Kronecker deltas:
\be \label{sdelta}
d_{a_1a_2 \ldots a_{2n-1} a_{2n}} = 
\delta_{(a_1 a_2} \delta_{a_3 a_4} \! \ldots \delta_{a_{2n-1} a_{2n})} . 
\ee
Such conservation laws hold in {\em any}
classically conformally-invariant theory. More interesting are the
equations
\be\label{curr}
\partial_- \tr (j_{+}^m) = 
\partial_+ \tr (j_{-}^m) = 0 \; 
\ee
which correspond to choosing 
\be \label{strace}
d_{a_1a_2 \ldots a_m} = s_{a_1a_2 \ldots a_m}:= 
\str(t^{a_1} t^{a_2} \! \ldots t^{a_m}) .
\ee
with `sTr' denoting the trace of a completely symmetrised product of 
matrices. For $su(N)$ these tensors exist for any positive integer $m$.
For $so(N)$ or $sp(N)$ on the other hand, they are non-trivial 
only when $m$ is an even integer, vanishing identically when $m$ is odd.

These observations lead us to a more detailed consideration of
invariant tensors. There are infinitely many invariant tensors for each 
algebra $\Lg$, but only $\rank(\Lg)$ independent or {\it primitive\/} 
$d$-tensors and Casimirs (see {\it e.g.}~\cite{azca97}), whose degrees
equal the exponents of ${\Lg}$ plus one. 
For future reference we list the exponents of each classical algebra,
together with the value of its Coxeter number, $h$ and the dimension of
its fundamental representation, $N$:
\begin{equation}
\label{exps}
\begin{array}{rcll}
a_\ell= su(\ell\!+\!1) & 1,2,3,\ldots,\ell & h=\ell\!+\!1&
\hspace{0.2in} N=\ell\!+\!1\\ 
b_\ell= so(2\ell\!+\!1) & 1,3,5,\ldots,2\ell\!-\!1 & h=2\ell&
\hspace{0.2in} N=2\ell\!+\!1\\
c_\ell= sp(2\ell) & 1,3,5,\ldots,2\ell\!-\!1  & h=2\ell& \hspace{0.2in}
N=2\ell\\
d_\ell= so(2\ell) & \hspace{0.2in}1,3,5,\ldots,2\ell\!-\!3\,;\,\ell\!-\!1 
\hspace{0.2in} &h=2\ell\!-\!2& \hspace{0.2in} N=2\ell
\end{array}
\end{equation}
All other invariant tensors 
can be expressed as polynomials in the primitive tensors 
and the structure constants 
$f_{abc}$. The choice of primitive tensors is certainly not unique.
In particular, we can modify any given choice
by adding terms involving {\it compound tensors} of the form 
$u_{(a_1 \ldots a_r} v_{b_1 \ldots b_s )}$.
For the classical algebras, the primitive tensors can
be chosen to be symmetrised traces $s_{a_1 \ldots a_m}$ as in 
(\ref{strace}), with one exception. 
This exception is the Pfaffian invariant for 
$so(2\ell)$, which has rank $\ell$ and which can be written
\be\label{pfaff}
d_{a_1...a_\ell} = p_{a_1...a_\ell}:= 
{1 \over 2^\ell \, \ell!} \, \epsilon_{i_1j_1 \ldots i_\ell j_\ell} 
\, (t^{a_1})_{i_1 j_1} \ldots
(t^{a_\ell})_{i_\ell j_\ell} \, .
\ee
Although the Pfaffian cannot itself be expressed as a polynomial in 
symmetrised traces, if $X$ is any element of the Lie algebra then 
it is always possible to express
$(p_{a_1 \ldots a_\ell} X^{a_1} \ldots X^{a_\ell})^2 = {\rm det} (X)$ 
as a polynomial in traces of powers of $X$.

We are interested in the behaviour of the general conserved charges 
\be
q_{\pm s} = \int \, d_{a_1 a_2 \ldots a_m}
j_\pm^{a_1} j_\pm^{a_2} \ldots j_\pm^{a_m} \, dx
\label{gencharge}
\ee
which we label by their spins $\pm s = \pm (m-1)$, {\it i.e.} their
eigenvalues under the Lorentz boost generator. 
In particular, the Poisson bracket algebra of these charges
can be calculated directly from (\ref{LCPBs}).
This was done in \cite{EHMM2} for the 
case $\lambda = 0$ (the PCM) and it was observed that 
the ultralocal terms in (\ref{LCPBs}) 
(those not involving $\delta'$) never contribute,
irrespective of their individual coefficients.
Now the effect of introducing a WZ term is evidently 
just a modification of these coefficients by some $\lambda$-dependent 
functions.
The arguments used in \cite{EHMM2} therefore suffice to show that
the ultralocal terms still do not contribute,
even in the more general case
$\lambda \neq 0$.
An immediate consequence is that charges (\ref{gencharge}) with 
spins of opposite sign always commute, since the brackets of
$j_+$ with $j_-$ involve {\it only} ultralocal terms.
For charges whose spins have the same sign, the situation is more 
complicated, since there may then be a 
contribution from the non-ultralocal term (proportional to $\delta'$) 
in (\ref{LCPBs}). For example, $\{ q_s , q_r \}$ is easily seen 
to be proportional to the integral of 
\be\label{pain}
d^{(s+1)}_{a_1 a_2 \ldots a_s c} d^{(r+1)}_{b_1 \ldots b_{r-1} e c}
j_+^{a_1} j^{a_2}_+ \ldots j^{a_s}_+
j_+^{b_1} \ldots j^{b_{r-1}}_+ \del_1 j^{e} \ .
\ee

Let us focus for definiteness on charges of positive spin and introduce 
the notation 
\be
\label{Jdef} 
{\cal J}_m = \tr (j_+^m) = 
s_{a_1 \ldots a_m} j^{a_1}_+ \ldots j^{a_m}_+
\, . 
\ee
The Poisson brackets of these currents are readily calculated 
\cite{balog90,EHMM2}, with the result 
\begin{eqnarray}
\{ \J_m (x) , \J_n (y) \} = - 2 mn \kappa (1{+}\lambda)^2 
\left [ \left(\J_{m+n-2}(x) - {1 \over N} \J_{m-1}(x) \, \J_{n-1}(x)
\right) \dpxy \right . \nonumber \\
\left . + {n{-}1 \over n{+}m{-}2} \, \J'_{m+n-2}(x) \, \dxy 
- {1 \over N} \, \J_{m-1} (x) \, \J'_{n-1}(x) \, \dxy 
\, \right ]
\label{JPBs}
\end{eqnarray}
This holds for each of the algebras $su(N)$, $so(N)$ and $sp(N)$,
though in the latter two cases the integers $m$ and $n$ must be 
taken to be even and the terms with coefficients $1/N$ then vanish.
By virtue of our earlier remarks concerning primitive invariant 
tensors, {\it any\/} positive-spin 
current of the general type (\ref{gencons}) 
can be expressed as a polynomial in a finite number of the 
currents $\J_m$, together with, for the case of $so(2\ell)$, the Pfaffian
current, which we write
\be\label{Pdef}
{\cal P}_\ell = 
p_{a_1 \ldots a_\ell} j^{a_1}_+ \ldots j^{a_\ell}_+ \ .
\ee
We remarked earlier that the Pfaffian can also be expressed in terms of
symmetrised traces, but only by taking a {\em square root} of a polynomial.
In practice such an expression may be rather inconvenient, although
for our purposes this will be sufficient.
Direct ways of calculating Poisson brackets involving the Pfaffian current 
are described in \cite{EHMM2}.

Our aim now is to identify certain natural families of local 
conserved charges of type (\ref{gencharge}) 
which all have vanishing Poisson brackets with one another.
For the orthogonal and symplectic algebras, for which the 
$1/N$ terms in (\ref{JPBs}) vanish, we see that the currents 
$\J_m$ already yield commuting charges.
This leaves out the case of $su(N)$, however, and also the Pfaffian
primitive invariant in $so(2 \ell)$. In the next section we 
shall give a universal formula which defines sets of commuting 
charges in any PCWZM based on a classical algebra.

To conclude this section, a number of comments are in order.
Firstly, we should mention in passing the existence of a much larger 
set of conserved local quantities in any PCWZM, consisting of  
{\it differential\/} polynomials in the currents (\ref{gencons}).
These will have no role to play in the remainder of this paper.
More importantly for us, 
there is a quite different but no less dramatic 
increase in the number of 
local charges for the special case of the WZW model. With $\lambda=+1$, say,
the Lie-algebra-valued current $j^a_+$ is itself holomorphic,
as in (\ref{wzw}). Consequently, the tensor $d$ appearing 
in (\ref{gencons}) need no longer be invariant in order to 
give rise to a holomorphic current:
{\it arbitrary\/} polynomials in the components $j^a_+$ are 
automatically holomorphic. We shall not investigate 
this larger set of currents in the WZW model in any detail, 
but knowledge of its existence will prove helpful later.

Finally, there is a related point concerning the quantum theory.
Away from the critical points $\lambda = \pm1$, the scale-invariance of 
the classical PCWZM will be broken quantum-mechanically \cite{WZW}.
We would therefore expect that the quantum versions of the classical
conservation laws (\ref{gencons}) are no longer of 
holomorphic form, but rather modified by anomalies (see
{\it e.g.} \cite{gold80}). Our detailed knowledge of the quantum
conservation laws is rather incomplete---see \cite{EHMM2} for a summary
of the PCM. At critical values of the coupling $\lambda = \pm1 $,
however, we obtain the quantum conformally-invariant WZW theories, and
we then expect the holomorphic form of the conservation equations to
persist quantum-mechanically.

\subsection{Commuting families of local charges}
\label{chargessect}

We now explain how the main results of \cite{EHMM2} regarding
commuting sets of local charges can be generalised to the PCWZMs.

For each of the classical algebras $a_\ell$, $b_\ell$, $c_\ell$ and
$d_\ell$, we introduce the generating functions $A(x,\mu)$ and
$F(x,\mu)$ by 
\be\label{Agen}
A (x , \mu) = \exp F(x, \mu) 
= \det ( 1 - \mu j_+ (x) ) 
\ee
so that 
\be\label{Fgen}
F(x , \mu) = \tr \log (1 - \mu j_+(x) ) 
= - \sum_{r=2}^\infty {\mu^r \over  r} \J_r (x) \; . 
\ee
Now define polynomials $\K_{s+1}$ in the currents $\J_m$, 
which have homogeneous spin 
$s{+}1$, by 
\be\label{kdef}
-{1 \over s{+}1} \, 
{\cal K}_{s+1} = A(x,\mu)^{s/h } \; \; {\Big |}_{ \mu^{s+1} } 
= \exp {s \over h } F(x, \mu) \; \; {\Big |}_{ \mu^{s+1} }  
\ee
where $h$ is the Coxeter number of the algebra, as given in (\ref{exps}).
Note that $A(x, \mu)$ is a polynomial in $\mu$ with degree equal to the 
dimension of the defining representation,
whereas $F(x,\mu)$ and fractional powers of $A(x,\mu)$ must be 
defined as power series in $\mu$ with infinitely many terms.
In defining ${\cal K}_{s+1}$ by extracting the indicated coefficient,
the generating function is to be expanded in ascending powers of $\mu$:
\be
A( x , \mu)^{s/h} = \sum_{r=0}^{\infty} \mu^r \, 
{\cal A}_r^{(s/h)} (x)
\qquad \Rightarrow \qquad
-{1 \over s{+}1} \,{\cal K}_{s+1} = {\cal A}^{(s/h)}_{s+1}.
\ee

The coefficient of the current in (\ref{kdef}) is 
chosen\footnote{In \cite{EHMM2} 
we adopted a number of different normalizations for these currents, 
as well as for the Pfaffian in (\ref{pfaff}) and (\ref{Pdef}). 
Such factors are obviously irrelevant to whether the corresponding 
charges commute, but they must be borne in mind when making comparisons 
with certain formulas in \cite{EHMM2}.}
to ensure that $\K_n$ has leading term $\J_n$.
The first few examples of these new currents are: 
\begin{eqnarray}
\K_2 & = & \J_2 
\nonumber\\
\K_3 & = & \J_3 
\nonumber\\
\K_4 & = & \J_4 - {3 \over 2 h} \, \J_2^2 
\nonumber\\
\K_5 & = & \J_5 - {10\over 3 h} \, \J_3\J_2 
\nonumber\\
\K_6 & = & \J_6 - {5\over 3 h} \, \J_3^2 
- {15\over 4 h} \J_4\J_2 + {25\over 8 h^2} \, \J_2^3 
\label{kexs}
\end{eqnarray}
These formulas apply to all algebras, though for the 
orthogonal and symplectic cases we must keep in mind that only the 
currents of even spin are non-vanishing.

The new currents 
${\cal K}_{s+1}$ are non-trivial 
precisely when $s$ is an exponent of the algebra $\Lg$ 
modulo its Coxeter number $h$.
If we consider $\Lg = su(N)$, the currents vanish when
$s/h$ is an integer, since $A(x,\mu)^{s/h}$ is then a polynomial in $\mu$
of degree $s$ (rather than a power series in $\mu$) 
and so the definition (\ref{kdef}) becomes empty.
For the orthogonal and symplectic algebras,
$h$ is always even, whereas 
the currents are non-vanishing only when $s$ is odd.
We have thus defined 
infinite sequences of currents or charges, each associated to a
primitive invariant tensor of type (\ref{strace}) of $\Lg$ and 
with spins repeating mod $h$ within each sequence.

The only primitive invariant missing from the discussion is 
the Pfaffian for $\Lg = so(2\ell)$.
To include this on the same footing, we must find an infinite
family of currents $\p_{\ell + ah}$ where $a = 0, 1, 2, \ldots$
and $h=2(\ell-1)$ is the Coxeter number of $so(2\ell)$.
It is not immediately obvious how this should be done,
but it turns out that this final sequence is already contained in the 
formulas given above, in the following rather surprising way.

First note that when $\Lg = so(2 \ell)$ the term of highest degree 
in $A(x,\mu)$ is $\mu^{2\ell} \p^2_\ell$. On extracting this factor
from $A(x,\mu)$, we are left with a polynomial in $1/\mu$, and this allows 
us to consider an expansion for 
$A(x,\mu)^{s/h}$ in decreasing, rather than increasing, powers of $\mu$.
Moreover, our earlier formula, on the right-hand-side of 
(\ref{kdef}), still makes perfect sense
with $h = 2(\ell -1)$ if we take
$s = (2a+1)(\ell -1)$ with 
$a$ a non-negative integer, and we can use it to define
\be\label{pfdef}
{\cal P}_{s+1} = A(x,\mu)^{s/h } 
\; \; {\Big |}_{ \mu^{s+1} } 
\ee
To be more explicit, the expansion in inverse powers of $\mu$ takes the form
\be 
A( x , \mu)^{s/h} = \mu^{(2a+1) \ell} \sum_{r=0}^{\infty} \mu^{-r} \, 
{\cal A}_{-r}^{(s/h)} (x)
\qquad \Rightarrow \qquad
{\cal P}_{\ell + a h} = {\cal A}^{(s/h)}_{-2a}
\ee
for $a= 0, 1, 2, \ldots \, \, $.
The first of these $(a=0)$ is indeed the Pfaffian,
and the following members of the sequence are its desired generalisations.

The importance of the new currents defined in (\ref{kdef}) and 
(\ref{pfdef}) is that their 
charges $\int \K_{s+1} \, dx$ and $\int \p_{s+1} \, dx$ 
all have vanishing Poisson brackets in the PCWZM.
This was proved for the case $\lambda =0$ in \cite{EHMM2}
by calculating the Poisson brackets of 
$A (x,\mu)^p$ and $A(y,\nu)^q$ and then showing that  
certain coefficients in the expansions were zero provided 
the powers $p$ and $q$ were chosen appropriately.
The extension from the PCM to the PCWZM is straightforward
once we know the effect of the WZ term, as given by the dependence on 
$\lambda$, in the brackets (\ref{LCPBs}) and (\ref{JPBs}).
Indeed, since $\lambda$ appears in (\ref{JPBs}) only through
the overall factor $(1+\lambda)^2$, it will appear in just 
the same way in brackets of $F$, $A$, powers of $A$, and hence 
in the brackets of the conserved charges.
Since changing zero by an overall factor still gives zero,
the calculations of \cite{EHMM2} 
imply that the charges commute in the PCWZM as well
as in the PCM.

There are actually more general sets of commuting charges for the 
algebras $so(2\ell +1)$ and $sp(2\ell)$, which correspond to replacing
$1/h$ by an arbitrary real number $\alpha$ in (\ref{kdef}).
The same freedom exists for the trace-type currents in the case of 
$so(2\ell)$, as far as their mutual commutation is concerned, but requiring
them to commute with the Pfaffian charges fixes $\alpha =1/h$.

The formula for the currents $\K_m$ in terms of $\J_m$ amounts to
a new choice $d_{a_1 \ldots a_m} = k_{a_1 \ldots a_m}$ 
in (\ref{gencons}); thus
\[
\K_m = k_{a_1 \ldots a_m} j^{a_1}_+ \ldots j^{a_m}_+ \ .
\]
Moreover the expressions for these new tensors in
terms of symmetric traces, of the form,
\[
k_{a_1 \ldots a_m}  = s_{a_1 \ldots a_m}  + ({\rm compound~terms})
\]
may be regarded as a new choice for the primitive 
symmetric invariants. In addition to
$k_{a_1 a_2} = s_{a_1 a_2}$ and 
$k_{a_1 a_2 a_3} = s_{a_1 a_2 a_3}$ we have the first few non-trivial
examples
\begin{eqnarray}
k_{a_1 a_2 a_3 a_4} & = & s_{a_1 a_2 a_3 a_4} 
- {3 \over 2 h} \, s_{(a_1 a_2} s_{a_3 a_4)}
\nonumber\\
k_{a_1 a_2 a_3 a_4 a_5} & = & s_{a_1 a_2 a_3 a_4 a_5} 
- {10\over 3 h} \, s_{(a_1 a_2 a_3} s_{a_4 a_5)}
\nonumber\\
k_{a_1 a_2 a_3 a_4 a_5 a_6} & = & s_{a_1 a_2 a_3 a_4 a_5 a_6} 
- {5\over 3 h} \, s_{(a_1 a_2 a_3} s_{a_4 a_5 a_6)} 
\nonumber \\
&&
- {15\over 4 h} s_{(a_1 a_2 a_3 a_4} s_{ a_5 a_6)} 
+ {25\over 8 h^2} \, s_{(a_1 a_2} s_{a_3 a_4} s_{ a_5 a_6)} 
\label{suN}
\end{eqnarray}
The vanishing of the charge Poisson brackets implies 
the algebraic property 
\be\label{keqn}
  {k^{(s+1)}_{(a_1 \ldots a_s}}^c k^{(r+1)}_{a_{s+1} \ldots a_{s+r-1})bc}
    = {k^{(s+1)}_{(a_1 \ldots a_s}}^c k^{(r+1)}_{a_{s+1} \ldots a_{s+r-1}b)c}
    \, ,
\ee
since this is a necessary and sufficient condition for 
the integrand in (\ref{pain}) to be a total derivative. 
This observation will prove useful later.

\subsection{Non-local charges}

In this section we construct the non-local charges in the 
PCWZM and then show that they 
commute with any charge arising from a local current (\ref{gencons}).

Recall that the PCM \cite{EHMM2} contains infinitely many
conserved, non-local charges, which generate a Yangian 
$Y(\Lg)$ \cite{mack92}. 
In fact there are two copies of this structure, 
constructed either from $j_\mu^L$ or $j_\mu^R$, 
and so the model has a charge algebra $Y_L(\Lg)\times Y_R(\Lg)$.
The charges are constructed \cite{brez79} (again for either $L$ or $R$; 
we specialise to $R$) from an infinite sequence of 
currents $j_\mu^{(r)}$, with 
$r=0, 1, 2, \ldots $, obeying 
\[
\del^\mu j_\mu^{(r)} = 0 
\quad \Leftrightarrow \quad
j^{(r)}_\pm = \pm \del_\pm \chi^{(r)}
\]
for some scalar functions $\chi^{(r)}$.
They are defined by
\[
j_\pm^{(0)} = j_\pm \, , 
\qquad 
j_\pm^{(1)} = (1\pm\lambda) 
(\, \partial_\pm \chi^{(0)} 
- {\textstyle {1 \over 2}}[E_\pm, \chi^{(0)} ] \, ) 
\]
and
\[ 
j_\mu^{(r+1)} = \nabla_\mu\chi^{(r)} =  
(1\pm\lambda) (\, \partial_\pm \chi^{(r)} - [E_\pm, \chi^{(r)} ] \, ) 
\, , \qquad r\geq1 ,
\]
which also defines the covariant derivative $\nabla_\mu$.
Note the factor of one half in the definition of 
$j_\mu^{(1)}$ which is needed to ensure that it is conserved.
Conservation of $j_\mu^{(r)}$ for $r > 1$ is  
easily established by induction, using the properties
$[\del_\mu , \nabla^\mu ] = [ \nabla_\mu , \nabla_\nu ] = 0$.
The first two conserved charges are\footnote{
Bernard \cite{Bern} writes down a closely related procedure
in which, using the freedom inherent in the Yangian, he effectively
subtracts $2\lambda Q^{(0)a}$ from $Q^{(1)a}$, so that in the conformal
limit $Q^{(1)a}$ is purely non-local. See also \cite{A,deV}.}
\begin{eqnarray*}
Q^{(0)a} & = & \int dx\, j^{a}_0(x)  \\
Q^{(1)a} & = & \int dx \, \left(\,
j^{a}_1(x) + \lambda j^{a}_0(x)
-{1\over 2\kappa}f^{abc}\int^x dy\,j^{b}_0(x)
j^{c}_0(y) \right) \,.
\end{eqnarray*}

We would expect $Q^{(0)}$ and $Q^{(1)}$ to form a Yangian, as in 
the $\lambda=0$ case, and the ultralocal parts of the Poisson brackets 
have the behaviour necessary for this. However, there is a problem
with the non-ultralocal terms (the derivatives of delta functions).
When $\lambda=0$ the only non-ultralocal term appears in the
$\{j_0,j_1\}$ bracket. Ambiguities in the charge brackets
can then be resolved by letting each charge be defined with
a range of integration from $-L$ to $L$, then letting 
$L\rightarrow\infty$. The brackets which define the Yangian 
are then independent of the order in which the limits $L_1\rightarrow\infty,
L_2\rightarrow\infty$ for the two charges are taken.
For $\lambda\neq0$ this is no longer possible. There are non-ultralocal
terms in each of the current brackets and these lead to an
ambiguity in $\{ Q^{(0)\,a},Q^{(1)\,b} \}$ which cannot be resolved, 
so that the Yangian is no longer well-defined -- an issue which remains 
to be understood.

Although the algebraic structure generated by $Q^{(0)b}$ and 
$Q^{(1)b}$ may be ambiguous when $\lambda \neq 0$, their brackets with 
a general local charge $q_s$ given by (\ref{gencharge}) 
are still well-defined.
It is straightforward to show that $\{q_s,Q^{(0)b}\}=0$.
In considering $\{q_s,Q^{(1)b}\}$, recall \cite{EHMM2} 
that in the $\lambda=0$ case the local and non-local contributions 
from $Q^{(1)b}$ conspired to cancel. The same now occurs with $\lambda\neq0$: 
the $\int j^{(0)b}_1$ terms yield extra contributions
\[
\int dx\, f^{a_1bc} (2\lambda j^{(0)c}_1 - \lambda^2j^{(0)c}_0)
d_{a_1a_2...a_{s+1}}j^{(0)a_2}_+j^{(0)a_3}_+ \ldots
j^{(0)a_{s+1}}_+\,,
\]
the $\lambda j^{(0)b}_0$ terms give nothing new, and the non-local
terms give extra contributions
\[
\int dx\,f^{a_1bc} (2\lambda j^{(0)c}_0 + \lambda^2j^{(0)c}_0)
d_{a_1a_2...a_{s+1}}j^{(0)a_2}_+j^{(0)a_3}_+ \ldots j^{(0)a_{s+1}}_+
\]
(we are assuming $s >0$).
The $\lambda^2$ terms cancel when these expressions are added, 
and the others sum to produce a factor $2\lambda j^{(0)c}_+$,
which gives zero on using invariance of $d$.

\section{The super PCM (and super WZW model)}

In the rest of this paper we shall consider supersymmetric extensions 
of the bosonic models discussed above. Supersymmetric principal chiral 
models (SPCMs) have long been known
to be integrable (see {\it e.g.} \cite{SPCM}) 
but our understanding of them is much less complete 
than for their bosonic counterparts. 
Indeed, an S-matrix for the $SU(N)$ SPCM has been 
conjectured (and tested) only fairly recently \cite{EH}, 
and it is far from clear how this particular construction could be 
generalised to other groups.
It is interesting therefore to consider whether a study of local charges, 
and the possible emergence of some version of Dorey's rule 
\cite{corri94,dorey91}, might be 
helpful for the eventual determination of these exact scattering theories.
The super WZW model by contrast has received considerable attention in the 
context of conformal field theory--see {\it e.g.} \cite{DKPR,SWZW}.

In the last section we showed that the addition of a WZ term
to the bosonic PCM ultimately has no effect 
on the existence of the conserved currents (\ref{gencons}) nor on the 
construction of commuting local charges based on the currents (\ref{kdef}).
Moreover, we found that the model displayed qualitatively different behaviour
only at the critical WZW point.
A similar picture emerges for the supersymmetric extensions of these models.
Because the formulas and calculations are considerably 
more complicated for the supersymmetric theories, we shall simplify
the presentation by concentrating 
for the most part on the SPCM, without a WZ term.
Where the presence of a WZ term becomes significant, however, 
we shall mention its effects explicitly. In particular,
some special features of the super WZW theory are 
important in clarifying the relationship between the bosonic and 
supersymmetric cases.

\subsection{Superspace and conserved currents}

To write down the SPCM in a manifestly supersymmetric way we shall 
use superspace, with coordinates $(x^\mu, \theta^+, \theta^-)$.
The additional fermionic coordinates $\theta^\pm$ 
are real Grassmann numbers, with supercharges 
$Q_\pm = \del_{\theta^\pm} + i\theta^\pm \del_\pm$ 
and supercovariant derivatives 
$D_\pm = \partial_{\theta^\pm} - i\theta^\pm\partial_\pm$.
Each index $\pm$ signifies one unit of Lorentz spin on a bosonic object, 
but a $1/2$-unit of spin on a fermionic object. Upper and lower indices
denote opposite Lorentz weights. (A fuller discussion 
is given in an appendix, section 7.)

In analysing the SPCM we will need to understand 
the implications of conservation equations in 
superspace, which have the general form  
\be\label{nonhol}
D_+J_- - D_-J_+ =0 \, .
\ee
Let us take the current components
$J_\pm$ to be fermionic superfields,
each carrying a single Lorentz spinor index, with all other possible 
internal or Lorentz indices suppressed
(more elaborate possibilities can be dealt with straightforwardly).
To examine the $x$-space content 
of the above equation we define the 
component expansions
\bea\nonumber
J_+ & = & \alpha_+ + \theta^+ j_+ + \theta^- u 
+ i \theta^+\theta^-\beta_+ \, , \\
J_- & = & \alpha_- + \theta^- j_- + \theta^+ v 
+ i \theta^-\theta^+\beta_- 
\label{cpts}\eea
in which all the fields are real, with $\alpha_\pm$ and $\beta_\pm$ fermionic, 
while $j_\pm$, $u$, $v$ are bosonic.
Now (\ref{nonhol}) is equivalent to
\bea
\partial_+ j_- + \partial_- j_+ & = & 0  \label{e1}\\ 
\del_- \alpha_+ &=& \beta_- \label{e2}\\
\del_+ \alpha_- &=& \beta_+ \label{e3}\\
u&=&v\label{e4} \,.
\eea
The first of these equations is the usual conservation equation for a bosonic
current with light-cone components $j_\pm$, 
and the corresponding conserved charge can be written
either as an integral in $x$-space or directly in superspace: 
\be\label{scharge}
B= \int (dx^+ j_+ - d x^- j_-)
= \int (dx^+ d \theta^+ J_+ - d x^- d \theta^- J_-) \, ,
\ee
where the $x$-integrals are understood to be taken over a space-like
curve. 

The remaining equations (\ref{e2})-(\ref{e4}) 
do not, in general, express any additional
conservation laws.
This is consistent
with the fact that the charge $B$ is always invariant under supersymmetry: 
$\delta_\epsilon B=0$.
This follows from either of the expressions in (\ref{scharge}), indeed,
it follows from the second expression simply because $J_\pm$ are
superfields. Thus it is not possible to discover a `superpartner
charge' to $B$ in this manner. (See also \cite{Ferr78}.)

Nevertheless, there are some important special circumstances in
which we know something more about the superspace current, thereby giving
extra content to (\ref{e2})-(\ref{e4}) and implying that there are 
additional conserved charges which are superpartners to $B$.
The simplest example 
is that of a {\it holomorphic\/} conservation law in
superspace, for which $J_- = 0$. In this case we clearly have 
\be \label{hol}
D_- J_+ = 0 \,  ,
\qquad \Rightarrow \qquad
J_+ = \alpha_+ + \theta^+ j_+ 
\qquad {\rm with} \qquad
\del_- j_+ = \del_- \alpha_+ = 0 \, .
\ee
Now in addition to the previous bosonic charge $B$ we also have a fermionic 
charge: 
\be \label{chargepair}
F = \int d x^+ \, \alpha_+ \, , \qquad B= \int d x^+ j_+ \,.
\ee
The action of supersymmetry on the currents is 
\be\label{susyhol}
\delta_\epsilon \alpha_+ = \epsilon^+ j_+ \, , \qquad
\delta_\epsilon j_+ = \epsilon^+ \del_+ \alpha_+
\ee
and so on the corresponding charges we have
\be\label{part} 
\delta_\epsilon F = \epsilon^+ B \, , \qquad \delta_\epsilon B = 0 \, .
\ee
Notice that our earlier conclusion regarding invariance of $B$ 
is unaltered.

These observations will be useful shortly.
Some more extensive comments about superspace conservation laws are
collected in an appendix, section 7.

\subsection{The SPCM lagrangian and symmetries}

To define the supersymmetric principal chiral model (SPCM) 
we introduce a superfield
$G(x , \theta)$ with values in ${\g}$.
The superspace lagrangian is 
\be 
{\cal L}= {1 \over 2} \tr( D_+ G^{-1} D_- G ) 
\ee
where we immediately set the overall coupling constant factor to unity.
This has a continuous symmetry
\be 
{\cal G}_L \times {\cal G}_R \; : \qquad G \mapsto U^{\vphantom{1}}_LGU_R^{-1}
\ee
and there are conserved, Lie algebra-valued superspace currents associated 
with each factor. As with the bosonic PCM, it will suffice to deal with just
one of these, which we choose to be the current 
corresponding to ${\cal G}_R$, namely,
\be 
J_\pm = - i G^{-1}D_\pm G 
\ee 
(the current for ${\cal G}_L$ is then $-G J_\pm G^{-1}$).
In addition to the superspace conservation equation (\ref{nonhol})
we have identically a zero-curvature condition 
in superspace:
\be \label{scurv}
D_+ J_- + D_- J_+ +  i \{ J_+ , J_- \} = 0 \, .
\ee
Combining these, the superspace equations of motion of the
SPCM can be written
\be\label{supercons}
D_+J_- = D_- J_+ = - {i\over 2} \{J_+, J_- \} \, .
\ee

To reveal the component ($x$-space) content of the super PCM we 
can expand
\be\label{Gcomp} 
G(x , \theta) = g(x) ( 1 + i \theta^+ \psi_+ (x) + i
\theta^- \psi_- (x) + i \theta^+ \theta^- \sigma (x) \, ) \; .
\ee
The fermions $\psi_\pm (x)$ take values in $\Lg$ and are the 
superpartners of the group-valued fields $g(x)$.
The ${\cal G}_L \times {\cal G}_R$ symmetry
acts on the component fields by 
$ g\mapsto U_L g U_R^{-1}$, and $\psi_\pm\mapsto U_R \psi_\pm U_R^{-1}$,
so that the fermions transform only under the ${\cal G}_R$ factor of the 
symmetry group.\footnote{
An alternative component expansion
$
G = \exp(i\theta^+\tilde{\psi}_+ + i \theta^-
\tilde{\psi}_- + i \theta^+\theta^-\tilde{\sigma}) g
$
would result in 
$
g\mapsto U_LgU_R^{-1}$ and 
$\tilde{\psi}_\pm\mapsto U_L
\tilde{\psi}_\pm U_L^{-1}
$.
The choices of fermions are of course related by 
$
\psi_\pm = g^{-1} \tilde{\psi}_\pm g 
$.}
The field $\sigma(x)$ turns out to be auxiliary, with an algebraic field 
equation. After its elimination, the final form of the component 
lagrangian is 
\begin{eqnarray}
L &=& -\half \, \tr\left( \, g^{-1}\partial_+g \, g^{-1}\partial_- g 
\, + \, i\psi_+ \partial_-\psi_+ \, + \, i \psi_- \partial_+ \psi_-
\right. \nonumber \\ 
& & 
\left. 
\, + \, \half i \psi_+ [g^{-1}\partial_- g,\psi_+] 
\, + \, \half i \psi_- [g^{-1}\partial_+ g,\psi_-]
\, + \, \half \psi_+^2\psi_-^2 \, \right) \, .
\end{eqnarray}

The component equations of motion which follow from this can 
equally-well be read off from the current conservation equations. 
We first calculate 
\begin{eqnarray}
J_+ = -i G^{-1}D_+G & =& 
\psi_+ -\theta^+(g^{-1}\partial_+g + i\psi_+^2)
- \half i \theta^- \{\psi_+,\psi_-\} \nonumber \\
&& -\;  i \theta^+\theta^-
(\, \partial_+\psi_- + [g^{-1}\partial_+g,\psi_-] + \half i 
[\psi_+^2,\psi_-] \, ) \label{Jpluscomp} \\
J_- = -iG^{-1}D_-G & =& 
\psi_- -\theta^-(g^{-1}\partial_-g + i\psi_-^2)
- \half i \theta^+ \{\psi_+,\psi_-\} \nonumber \\
&& -\;  i \theta^-\theta^+
(\, \partial_-\psi_+ + [g^{-1}\partial_-g,\psi_+] + \half i 
[\psi_-^2,\psi_+] \, ) \,  \label{Jminuscomp}
\end{eqnarray}
(having already eliminated the auxiliary field).  
We can then compare these component expansions to
(\ref{cpts}), and we find that the 
equations (\ref{e1}-\ref{e4}) 
imply that the bosonic current
\be\label{boscurr}
j_\pm = -(g^{-1}\partial_\pm g + i\psi_\pm^2)
\ee
is conserved, while the fermion equations of motion are
\be\label{dirac} 
\partial_\mp \psi_\pm +{1\over 2}[g^{-1}\partial_\mp g, \psi_\pm] 
+{i\over 4}[\psi_\mp^2,\psi_\pm] = 0 \,.
\ee 

As in the bosonic case, there can be important discrete symmetries of the 
SPCM. In particular, we have a ${\cal G}$-parity symmetry
\[ \pi: G \mapsto G^{-1} \qquad \Rightarrow \qquad
J_\pm \mapsto - G J_\pm G^{-1} \ .
\]
The derivatives $D_\pm J_\pm$ do not have definite behaviour under $\pi$,
so we introduce instead the combinations
\[ 
J_{\pm\pm} = D_\pm J_\pm + i J_\pm^2 \qquad \Rightarrow \qquad
J_{\pm\pm} \mapsto - G J_{\pm\pm}G^{-1} \ .
\] 
Similar modifications can be made to higher derivatives 
(which proved useful in \cite{EHMM1}).

The classical super PCM is superconformally invariant, with the 
non-vanishing components of the super energy-momentum tensor
obeying 
\be\label{holsem}
D_-\tr (J_+ J_{++}) =
D_+\tr (J_- J_{--}) =0 \ .
\ee
When expanded in components this contains conservation equations for 
both the supersymmetry current and the conventional (bosonic) 
energy momentum tensor.

\subsection{Poisson brackets in the SPCM}

Poisson brackets will always be written $\{ A , B\}$ but must 
be understood to be graded, {\it i.e.} 
antisymmetric if either $A$ or $B$ is bosonic, but symmetric if
both $A$ and $B$ are fermionic. They obey the Leibnitz rules
\[ 
\{ A , BC \} = \{ A , B\} C \pm B \{ A, C \}
\, , \qquad 
 \{ CB , A \} = C \{ B , A\} \pm \{ C, A \} B
\]
where the minus signs occur if and only if 
both $A$ and $B$ are fermionic.

The brackets for the bosonic conserved currents (\ref{boscurr}) are 
\begin{eqnarray}
\{j_0^a(x),j_0^b(y)\} & =  & f^{abc}j_0^c(x) \dxy 
\nonumber \\
\{j_0^a(x),j_1^b(y)\} & =  & f^{abc}j_1^c(x) \dxy + 
\delta^{ab} \dpxy \label{spcmpbs} \\
\{j_1^a(x),j_1^b(y)\} & =  & - \quarter i f^{abc} (\, h^c_+(x) + h^c_-(x) \, )
\dxy
\nonumber
\end{eqnarray}
or, in light-cone coordinates,
\begin{eqnarray}
\{j_\pm^a(x),j_\pm^b(y)\} & =  & \half f^{abc}\left(
3 j_\pm^c(x) - j_\mp^c(x) - 
\half i h^c_+ (x) - \half i h^c_- (x) 
\right) \dxy
\nonumber \\ 
&&\pm \, 2 \delta^{ab} \delta'(x{-}y) \\
\{j_+^a(x),j_-^b(y)\} & =  & \half  f^{abc}\left( 
j_+^c(x)  + j_-^c(x)  + \half i h^c_+ (x) + \half i h^c_- (x) 
\right)
\dxy \nonumber
\end{eqnarray}
where we have introduced the bosonic quantities 
\be
h_\pm = \psi_\pm^2 \ , \qquad h_\pm^a 
= \half \, f^{abc} \psi_\pm^b \psi_\pm^c \ .
\ee
The fermions obey
\be
\{ \psi^a_\pm (x) , \psi^b_\pm (y) \} = -i \delta^{ab} \delta(x{-}y)
\, , \qquad
\{ \psi^a_+ (x) , \psi^b_- (y) \} = 0 \ .
\ee
It is also useful to note that
\begin{eqnarray}\label{fermpbs}
\{ h^a_\pm (x) , \psi^b_\pm (y) \} & = & i f^{abc} \psi^c_\pm (x) \delta(x{-}y)
\\
\{ h^a_\pm (x) , h^b_\pm (y) \} & = & i f^{abc} h^c_\pm (x) \delta(x{-}y)
\end{eqnarray} 
Finally, there are non-trivial brackets between the bosonic currents 
and the fermions
\be
\{ j^a_\pm (x) , \psi^b_\pm (y) \} = 
\threehalf f^{abc} \psi^c_\pm (x) \dxy
\ , \qquad
\{ j^a_\pm (x) , \psi^b_\mp (y) \} = \half f^{abc} \psi^c_\mp (x) \dxy
\ee
A derivation of the brackets above is sketched in an appendix, section 6.

\subsection{Local conserved charges}

The simplest local conserved currents in the bosonic PCM are 
powers of the energy-momentum tensor (\ref{confpower}).
The super energy-momentum tensor in the SPCM is a fermionic 
quantity, however,
so we cannot take powers of it to obtain new conservation laws in
quite the same way. Let us therefore turn directly to the
generalisations of (\ref{curr}) and (\ref{gencons}).
The currents (\ref{curr}) in the bosonic PCM
can be generalised to the supersymmetric PCM in two ways.
First, we have\footnote{We restrict immediately to positive spins; there
are obviously analogous negative-spin currents annihilated by
$D_+$. See also \cite{HP92} for a related construction.}  
\be\label{oddcurr}
D_- \tr(J_+^{2n+1}) = 0 
\ee
which is odd under the discrete symmetry $\pi$.
The power of $J_+$ must be an odd integer, otherwise the expression 
would vanish identically, by Fermi statistics.
Second, we have 
\be \label{evencurr}
D_- \tr(J_+^{2n-1} J^{\phantom{n}}_{++}) = 
0  
\ee
which is even under $\pi$. 
The power of $J_+$ must again be odd, this time to prevent the expression
being a total $D_+$ derivative and hence giving a trivial conservation
equation. The first member of this sequence, with $n=1$, is the 
super-energy-momentum tensor.
Both (\ref{oddcurr}) and (\ref{evencurr}) 
follow directly from the superspace equations of motion 
(\ref{supercons}).

As in the bosonic case, one can generalise
these conservation equations by re-writing them in terms of
invariant tensors. Equation (\ref{oddcurr}) becomes 
\be\label{oddgen}
D_- ( \, \Omega_{a_1 a_2 \ldots a_{2n+1} } J^{a_1}_+ J^{a_2}_+ 
\ldots J^{a_{2n+1}}_+ \, ) = 0
\ee
where we define, from a symmetric $d$-tensor of rank $n{+}1$, a
completely antisymmetric tensor of rank $2n{+}1$ by
\be\label{omega} 
\Omega_{a_1 a_2 \ldots a_{2n+1}} =  
{1\over 2^n}
f_{ [ a_1 a_2}{}^{b_1} \ldots f_{a_{2n-1}
a_{2n}}{}^{b_n} d^{\,b_1 \ldots b_n}{}_{a_{2n+1} ]}  \ .
\ee
In a similar fashion, the second kind of conservation equation 
(\ref{evencurr}) becomes 
\be\label{evengen}
D_- ( \, \Lambda_{a_1 \ldots a_{2n-1} a_{2n} } 
J_+^{a_1} \ldots J^{a_{2n-1}}_{+} J^{a_{2n}}_{++} \, ) =
0
\ee
where now the relevant invariant tensor is even-rank,
\be\label{lambda}
\Lambda_{a_1 a_2 \ldots a_{2n-1} a_{2n} }
= 
{1\over 2^{n-1}}
f_{ [ a_1 a_2}{}^{b_1} \ldots f_{a_{2n-3}
a_{2n-2}}{}^{b_{n-1}} d^{b_1 \ldots b_{n-1}}{}_{a_{2n-1} ] a_{2n}} 
\ .
\ee
It has a more complicated structure in that it is antisymmetric only
on its first $2n{-}1$ indices.

It is clear that we need invariant tensors which are 
{\em antisymmetric\/} in some number of indices, in order to combine 
the fermionic currents $J_+$ into holomorphic expressions. 
An immediate consequence of this fact, however, is that there are only 
finitely many such expressions, in contrast to the infinitely many 
holomorphic currents (\ref{gencons}) in the bosonic PCM, which are based on 
{\em symmetric\/} invariant tensors.

It is useful to pin-point the precise algebraic properties of the 
$\Omega$ and $\Lambda$ tensors which are relevant here. It can be shown that
$\Omega$ vanishes whenever the symmetric tensor $d$ in (\ref{omega}) is of 
compound type. As a result, it is only the primitive part of $d$ which
contributes to the expression for $\Omega$ and moreover $\Omega$ is
independent (up to an overall factor) of how this primitive tensor is
chosen. The situation for $\Lambda$ is only slightly more
complicated. If $d$ is a compound tensor made up of just two primitive
tensors, then $\Lambda$ does not vanish, but it reduces to a product of
two $\Omega$ tensors. If $d$ is compound and made up of three or more
primitive tensors, then $\Lambda$ vanishes. These properties are
explained in detail in an appendix, section 9. 

To gain a better understanding of the superspace 
conservation equations, we expand them in component
fields, using (\ref{Jpluscomp}).
Both kinds of conserved current are holomorphic, and we distinguish 
their components with superscripts $\pm$ to indicate their behaviour 
under the $\g$-parity symmetry $\pi$. Thus 
\be
\Omega_{a_1 a_2 \ldots a_{2n+1} } J^{a_1}_+ J^{a_2}_+ 
\ldots J^{a_{2n+1}}_+ 
= \F^-_{n+{1\over2}} + (2n{+}1) \theta^+ \B^-_{n+1}
\ee
(the insertion of the factor $(2n{+}1)$ proves convenient) where
\begin{eqnarray} 
{\cal F}^-_{n+{1\over2}} 
& = & 
\Omega_{a_1 a_2 \ldots a_{2n+1} } \psi^{a_1}_+ \psi^{a_2}_+ 
\ldots \psi^{a_{2n+1}}_+
\nonumber\\
& = & 
d_{a_1 a_2 \ldots a_n a_{n+1} } h^{a_1}_+ h^{a_2}_+ \ldots
h_+^{a_n} \psi_+^{a_{n+1}} 
\label{oddfcomp}\\[2pt]
\B^-_{n+1} & = & 
\Omega_{a_1 \ldots a_{2n} a_{2n+1} } \psi^{a_1}_+ \ldots \psi^{a_{2n}}_+ 
j^{a_{2n+1}}_+
\nonumber\\
& = &
d_{a_1 a_2 \ldots a_{n} a_{n+1} } 
h^{a_1}_+ h^{a_2}_+ \ldots h_+^{a_n} j_+^{a_{n+1}} 
\label{oddbcomp} 
\end{eqnarray}
The expansion of the other currents is more complicated:
\be 
\Lambda_{a_1 \ldots a_{2n-1} a_{2n} } 
J_+^{a_1} \ldots J^{a_{2n-1}}_{+} J^{a_{2n}}_{++} 
= \F^+_{n+{1\over2}} \, + \, \theta^+ \B^+_{n+1}
\ee
where 
\vfill \eject
\begin{eqnarray}
{\cal F}^+_{n+{1\over2}} & = & \Lambda_{a_1 a_2 \ldots a_{2n-1} a_{2n} }
\psi^{a_1} \ldots \psi^{a_{2n-1}} j_+^{a_{2n}}
- \half i \F^-_{n+{1\over2}}
\nonumber \\
& = &
d_{a_1 a_2 a_3 \ldots a_{n+1} } j_+^{a_1} \psi_+^{a_2} h_+^{a_3} \ldots 
h_+^{a_{n+1}} - \half i \F^-_{n+{1\over2}},
\label{evenfcomp} \\[2pt]
{\cal B}^+_{n+1} & = & 
\Lambda_{a_1 a_2 \ldots a_{2n-1} a_{2n}} 
( \, (2n{-}1)j_+^{a_1} j^{a_{2n}}_+ + i \psi^{a_1}_+ \del_+
\psi^{a_{2n}}_+ \, )  \psi^{a_2}_+ \ldots \psi_+^{a_{2n-1}} 
- (n{-}1) i \B^-_{n+1}
\nonumber \\
& = & 
d_{a_1 a_2 a_3 \ldots a_{n+1}} 
( \, nj_+^{a_1} j^{a_2}_+ + i \psi^{a_1}_+ \del_+
\psi^{a_2}_+ \, )  h_+^{a_3} \ldots h_+^{a_{n+1}} - (n{-}1) i \B^-_{n+1}
\label{evenbcomp}
\end{eqnarray} 
(recall the bosonic quantity 
$h_+^a = \half f^{abc} \psi_+^b \psi_+^c$).

In either family of conservation laws, the fermionic and 
bosonic currents have spins $n{+}{1\over2}$ and $n{+}1$ respectively,
and so the corresponding conserved charges have spins
$n{-}{1\over2}$ and $n$ respectively. 
The fact that the currents are based on a primitive $d$-tensor of 
rank $n{+}1$ then implies that the values of $n$ are precisely the 
exponents of the algebra. 

Little can be said at present about the effects of quantisation on
these conservation equations, although counting arguments can be used to
demonstrate the persistence of some of them \cite{EHMM1}, following the 
methods of \cite{gold80,clark81}.
What is certain, however, is that (super)conformal invariance 
will be broken in passing from the classical to the quantum SPCM, 
and so one cannot expect the conservation equations to survive in 
(super)holomorphic form. 
Now we have already seen that a typical {\em non}-holomorphic 
superspace conservation equation (\ref{nonhol}) leads to a conserved 
bosonic charge without, in general, any fermionic partner.
This suggests that while the fermionic currents $\F^\pm$ are 
relevant to understanding the classical structure of the 
SPCM, it is the bosonic currents $\B^\pm$ which are of more central
importance in the quantum theory.

A last word of caution should be added.
It is quite possible for 
a pair of bosonic and fermionic charges to arise from a 
non-holomorphic conservation law, 
and a familiar example is provided by energy-momentum and its fermionic 
partner, supersymmetry.
The classical holomorphic equation (\ref{holsem}) 
for the super-energy-momentum tensor will certainly receive quantum 
modifications, and yet we expect that supersymmetry will also survive in 
the quantum theory.
Such behaviour is only possible when the non-holomorphic 
superspace current has a very particular structure.
For the case of the super-energy-momentum tensor, this special
structure is easily understood (as a consequence of Noether's Theorem)
and this is explained in an appendix, section 7.
We can see no reason to expect similar behaviour for the 
superspace conservation equations of higher spin.

\subsection{Comparing the bosonic and supersymmetric PCMs}

Essentially by construction, the super PCM reduces to the 
bosonic PCM when all fermions are set to zero.
The formulas for the conserved currents such as (\ref{curr}) and 
(\ref{oddcurr}) seem very similar superficially, but 
here the superspace notation and the fermionic character of the current 
in the SPCM hide some profound differences.
In the bosonic PCM there are infinitely many holomorphic
quantities (\ref{gencons}).
In the supersymmetric PCM we found two distinct sets (\ref{oddgen}) and 
(\ref{evengen}), each of them 
containing only finitely many holomorphic quantities.
Notice also that when we reduce the SPCM to the bosonic PCM by setting the 
fermions to zero, {\it all} the currents 
$\B^\pm$ and $\F^\pm$ vanish, with the exception of $\B^+_2$ (which 
is of course the $++$ component of the energy-momentum tensor).
This begs a question: Are there other 
conserved quantities in the SPCM which will  
reduce to any of the currents in (\ref{gencons}) when the fermions 
vanish?
We will argue that there are not, but to do so we must consider 
what possible forms such conservation equations might take, so as to 
be able to phrase the question more precisely.

First observe that the series of 
holomorphic quantities $\tr (j_+^m)$ in the bosonic PCM 
relies on the form of the equation of motion of the 
current: $\del_- j_+ = (const.)[ j_+ , j_-]$.
It is actually irrelevant for this purpose that $j_\pm$ is 
conserved. All that is important is that $\del_- j_+$ 
can be expressed as a commutator of $j_+$ with some other field in the 
Lie algebra.
In the SPCM, the bosonic current $j_\pm$ is defined by (\ref{boscurr}).
Any quantity of the form $k_+ = j_+ + \alpha \psi_+^2$ will reduce
to the conserved current component in the bosonic PCM when the
fermions are set to zero, and this is obviously the most general 
such polynomial in the current and the fermions with spin 1.
A precise way to pose the question of the 
last paragraph is now to ask whether it is possible to find 
$k_- = j_- + \beta \psi_-^2$ such that
\be\label{nogo}
\del_- k_+ = \gamma [ k_- , k_+ ]
\ee
in the SPCM.
This would be enough to ensure the existence of holomorphic 
quantities $\tr (k_+^m)$ in the SPCM of the type we seek.
After a short calculation, however, one finds that such an equation
implies an over-determined set of relations amongst the constants 
$\alpha$, $\beta$ and $\gamma$, and there is no solution.
Details of a more general calculation which includes the effects of a WZ 
term are given in the next section.

One might object that the form assumed in (\ref{nogo}) 
is already too restrictive.
We can avoid this assumption and still carry out a similar argument
if we focus on the simplest example of currents of spin 3 (in the 
SPCM based on $su(N )$ for example).
Thus, we can consider the most general $(\g_L \times \g_R)$-invariant 
polynomial in the fields $j_+$ and $\psi_+$ with this spin, and ask 
whether it can ever be holomorphic. It actually suffices to consider 
\be\label{holnogo}
\tr (j_+^3)  + \alpha \tr (j_+^2 \psi_+^2)
\ee
since the only other possible terms are
$\tr (j_+ \psi_+ j_+ \psi_+ ) = 0$ identically, by Fermi 
statistics and cyclicity of the trace, and 
$\tr (j_+ \psi_+^4)$, which we already know to be holomorphic, and which 
is therefore of no help in achieving our goal. 
Once again an explicit calculation reveals that there is no 
choice of $\alpha$ for which the expression above is holomorphic
(see also the following section).

These results may seem unexpected, but in fact
similar behaviour is known to occur in other integrable
models and their supersymmetric extensions. Specifically, 
for bosonic conformal Toda theories, and their
(1,0) supersymmetric extensions \cite{Pap}, one finds \cite{EM} that 
the higher-spin conserved quantities generating the ${\cal W}$-algebra of 
the bosonic theory have no generalisations (in a suitably precise sense, 
similar to the ones above).
Here, as there, it would appear that an `interference' arises between 
supersymmetry and the existence of currents with non-trivial spin.
The ultimate implications for these models are rather different 
however. In the (1,0) Toda theories, no other 
conserved quantities are known, and it was conjectured on this basis 
that these supersymmetric extensions are not integrable \cite{EM}.
In the SPCM there {\it are\/} higher-spin local (as well as non-local)
conserved quantities which guarantee integrability.
As we have seen, these are genuinely new features of the SPCM 
in the sense that they do not reduce to the 
familiar conserved quantities of the bosonic PCM when the fermions are 
set to zero. 

\subsection{The effect of a WZ term and the special nature of the WZW points}

So far we have dealt exclusively with the SPCM, allowing us to keep the 
presentation as simple as possible.
We now discuss briefly the modifications which arise on adding a WZ term,
which can easily be constructed in superspace, as in {\em e.g.\ }\cite{DKPR}.
The effect of a suitably normalised WZ term with coefficient $\lambda$ 
is to modify the equation 
obeyed by the currents $J_\pm = -i G^{-1} D_\pm G$ so that it becomes
\be\label{nonchS}
(1-\lambda) D_+ J_- - (1+\lambda) D_- J_+ = 0 \ .
\ee
Adopting the same component field definitions as before,
we find, after eliminating the auxiliary field,
\be\label{wzcurr}
( 1 + \lambda) \del_- j_+ 
+
( 1 - \lambda) \del_+ j_- 
= 0 \qquad {\rm where} \qquad j_\pm = - g^{-1} \del_\pm g - i \psi^2_\pm
\ee
and
\be\label{wzdirac} 
\partial_\mp \psi_\pm +{1\over 2}(1 \mp \lambda) 
[g^{-1}\partial_\mp g, \psi_\pm] 
+{i\over 4}(1 - \lambda^2)[\psi_\mp^2,\psi_\pm] = 0 \,.
\ee 
The construction of the two sets of superfield currents in (\ref{oddgen})
and (\ref{evengen}) is completely unaffected, no matter what the value 
of $\lambda$.

Some important qualitative differences 
arise at the critical values of the coupling $\lambda = \pm 1$ which define 
super WZW theories.
For $\lambda = 1 $, for instance, the current 
becomes super-holomorphic: $D_- J_+ = 0$.
This means of course that there is a 
simplified superfield expansion $J_+ = \psi_+ + \theta^+ j_+$ with
$ \del_- \psi_+ = \del_- j_+ = 0$.
These holomorphic equations of motion arise in conjunction with the 
super Kac-Moody symmetry of the critical theory \cite{DKPR,SWZW}.

Earlier we stated that there was no natural way to generalise 
the local conservation laws of the bosonic PCM to the SPCM by 
using an equation such as (\ref{nogo}), or by looking for a holomorphic 
quantity such as (\ref{holnogo}). 
These same calculations can be carried out more generally in the 
PCWZM and its supersymmetric counterpart, and it is instructive to 
elaborate on the details. 

First we use (\ref{wzcurr}) and (\ref{wzdirac}) to find 
\be
\del_- j_+ = -{1\over 2} (1{-}\lambda) [ j_+ , j_- ] 
-{i \over 4} (1{-}\lambda)^2 [ j_+ , \psi_-^2 ]
+{i \over 4} (1{-}\lambda^2) [ j_- , \psi_+^2 ]
+{1 \over 4} (1 {-} \lambda) (1 {-} \lambda^2) [ \psi^2_+ , \psi_-^2 ]
\label{wzeqn}
\ee
Now we apply this expression together with (\ref{wzdirac}) to calculate
both sides of 
\be \label{newnogo} 
\del_- k_+ = \gamma [ k_- , k_+ ] \qquad {\rm where} \qquad
k_+ = j_+ + i \alpha \psi_+^2 , \qquad
k_- = j_- + i \beta \psi_-^2 \ .
\ee 
Comparing coefficients of like terms on 
the left- and right-hand sides gives the relations
\begin{eqnarray}
\gamma & = & \half (1 - \lambda) \nonumber \\
\beta \gamma  & = & \quarter (1 - \lambda)^2 \nonumber \\
\alpha \gamma & = & \half \alpha (1 - \lambda) + \quarter (1 - \lambda^2)
\nonumber \\
\alpha \beta \gamma & = & \half \alpha (1 - \lambda) + 
\quarter (1 - \lambda^2 ) (1 - \lambda -\alpha)
\nonumber
\end{eqnarray}
Substituting for $\gamma$ from the first equation into the third 
reveals that the equations are consistent only for $\lambda^2= 1$,
the super WZW points. Thus if $\lambda \neq \pm 1$, including
the super PCM with $\lambda = 0$, then there is no relation 
of the type we seek.

If $\lambda =1$, the general solution is $\gamma = 0$, 
with $\alpha, \beta$ arbitrary. We then recover from (\ref{newnogo}) 
the familiar conditions $\del_- j_+ = \del_- \psi_+ = 0$.
If $\lambda = - 1$, the general solution is $\gamma = \beta =1$ and 
$\alpha$ arbitrary. This is also as expected, because 
at this second WZW point it is the left-transforming quantities
$g j_+ g^{-1}$, and $g \psi_+ g^{-1}$ which should be holomorphic,
and thus each of them should satisfy 
$\del_- (g X g^{-1}) = 0$, or equivalently, 
$\del_- X + [ g^{-1} \del_- g , X ] = 0$,
which is indeed the content of (\ref{newnogo}).
The solutions for $\lambda = \pm 1$ are therefore less
surprising than the lack of solutions for $\lambda \neq \pm1$.

In a similar fashion, we can examine the possibility of a holomorphic
quantity of the form (\ref{holnogo}). 
Using (\ref{wzeqn}) and (\ref{wzdirac}) it is a simple matter to calculate
$\del_- {\rm Tr} (j_+^3)$ and $\del_- {\rm Tr}(j_+^2 \psi_+^2)$.
Both expressions have an overall factor of $(\lambda^2 -1)$,
and the coefficients of like terms cannot be matched 
except when this vanishes. Thus, supersymmetry does not allow a 
generalisation of such a spin-3 current,
except in the super WZW models.

We drew attention previously to the fact that the bosonic WZW theory
(with $\lambda =1$ say) contains an enlarged set of holomorphic currents 
polynomial in $j_+^a$.
Similarly, for the super WZW theory (with $\lambda =1$)
the currents (\ref{oddgen}) and (\ref{evengen}) 
exist within a much larger set of holomorphic quantities 
consisting of arbitrary polynomials in both $j^a_+$ and $\psi^a_+$.
The holomorphic 
currents in the bosonic WZW theory obviously extend 
immediately to the super WZW theory.
This simple relationship between these enlarged sets of conserved 
quantities in the critical theories is of course responsible for 
the solutions to (\ref{newnogo}) found at $\lambda = \pm1$.
By contrast, our results indicate that there is no 
simple connection between the conserved quantities in 
the PCWZM and its super-extension at non-critical coupling 
$\lambda \neq \pm1$.

\subsection{Non-local charges}

The non-local charges for a supersymmetric sigma model may
be constructed \cite{curt80} in component formalism using a rather
complicated generalisation of the iterative procedure described
earlier for the bosonic case. The result is again $Y_L\times Y_R$:
there are no new fermionic superpartners for the charges.
The construction looks neater in the superfield formalism
\cite{chau86}, where it is a natural extension of the bosonic case,
and the lack of superpartners is accounted for by the general discussion we
gave in section 3. Here we include the effect of a superspace WZ term,
$\lambda\neq 0$.

We define a superspace connection acting on any
bosonic quantity $X$ in the Lie algebra (if
$X$ is fermionic we replace the commutator with an anti-commutator):
\[
\nabla_\pm X = (1\pm\lambda)\left(D_\pm X \,+\, i [ J_\pm,
X ] \right)
\quad \Rightarrow
\quad \{\nabla_+ ,\nabla_-\}=0 \;,
\]
by virtue of (\ref{scurv}).
{}From this we can define an infinite family of superfield
currents $J_\pm^{(r)}$ for $r=0,1,2\ldots$ which will
be conserved: 
\be\label{sbizz} 
D_-J_+^{(r)} - D_+J_-^{(r)} = 0 \quad \Leftrightarrow
\quad J_\pm^{(r)}= \pm D_\pm X^{(r)},
\ee
for some scalar superfields $X^{(r)}$.
The first two currents are 
\be
J_\pm^{(0)} = (1\pm\lambda)J_\pm \, ,
\qquad 
J_\pm^{(1)} = (1\pm\lambda) ( \, D_\pm X^{(0)}
- {\textstyle {1 \over 2}}[J_\pm , X^{(0)}] \, )
\ee
whose conservation follows from (\ref{nonchS}) and (\ref{scurv}).
The remaining currents are defined by (\ref{sbizz})
and
\be
J_\pm^{(r)} = \nabla_\pm X^{(r-1)} \, \qquad r > 1. 
\ee
It is easy to prove by induction that these are conserved:
if this holds for all $r \leq n$ then
\beaa
D_-J_+^{(n+1)} - D_+J_-^{(n+1)} 
&=& (D_-\nabla_+ - D_+\nabla_-)X^{(n)} \\
&=& - (\nabla_+D_- - \nabla_-D_+)X^{(n)} \\
&=& \nabla_+ J_-^{(n)}+\nabla_- J_+^{(n)} \\
&=& \{\nabla_+,\nabla_-\} X^{(n-1)} \qquad (n > 1) \\
&=&0 \,.
\eeaa

The corresponding non-local conserved charges are given by
\beaa
Q^{(n)} & = & \int(dx^+\,d\theta_+\,J_+^{(n)} \; - \; dx^- d\theta_- 
\,J_-^{(n)}) \\
&=& \int(dx^+\,j_+^{(n)} \;-\;dx^-\,j_-^{(n)}) \
 \hsp+\hsp \theta^-\int dx^+\partial_+\alpha_-^{(n)}
 \hsp-\hsp \theta^+\int dx^-\partial_-\alpha_+^{(n)} 
\eeaa
and it may be checked that $\alpha_\pm^{(n)} \rightarrow 0$
as $x\rightarrow\pm\infty$. 
(Our notation for current components was introduced in (\ref{cpts}).) 
The first two examples are
\begin{eqnarray}\label{snlcs}
Q^{(0)a} & = & \int dx\, j^{(0)a}_0(x) \\
Q^{(1)a} & = & \int dx \, \left(\,
j^{(0)a}_1(x) +  \lambda j^{(0)a}_0(x)  
+{i\over 2}\left((1{-}\lambda)^2h_-^a  -(1{+}\lambda)^2h_+^a\right)
\right. \nonumber\\
&& \qquad \left. -{1\over 2}f^{abc}\int^x dy\,j^{(0)b}_0(x)
j^{(0)c}_0(y) \right) . \nonumber
\end{eqnarray}
When $\lambda=0$ we make the link with 
the first paper of \cite{curt80} by pointing out that, comparing
with their eqns.~(2.12, 3.9), their
$B_\pm$ equals our ${1\over 2}i h_\pm$.

When $\lambda=0$
it is a straightforward though cumbersome calculation to show
that the non-local charges defined above commute with all four sets 
of local charges
obtained from (\ref{oddfcomp}-\ref{evenbcomp}); see the appendix, section 8. 
We have not attempted this calculation for $\lambda\neq0$.

\section{Classical current algebra and commuting local charges in the SPCM}

In the last section we constructed local, holomorphic superspace currents
(\ref{oddgen}) and (\ref{evengen})  
in the classical SPCM. Each such current 
gave a pair of fermionic and bosonic holomorphic currents
$\F^\pm_{n+1/2}$ and $\B^\pm_{n+1}$ in ordinary space.
Our aim now is to analyse and understand the properties of these 
currents at a level comparable to our treatment of the bosonic PCWZM in 
section 2. Thus our first goal will be to compute the Poisson bracket (PB)
algebra of a certain class of currents.
Using these results, we will then search for commuting sets 
of charges.

To carry out the Poisson bracket calculations, 
it proves convenient to introduce the modified currents
\be
\jh_+ = j_+ + \threehalf \,  i h_+ \ ,
\qquad
\jh_- = j_- + \half \, ih_+ + \half \, i h_-
\ee
By construction, these 
have vanishing PBs with the fermions $\psi_+$.
They can be shown to obey 
\begin{eqnarray}
\{\jh_+^a(x),\jh_+^b(y)\} & =  & \half f^{abc} \left(
3 \jh_+^c(x) - \jh_-^c(x) 
\right) \delta(x{-}y)
+ 2 \, \delta^{ab} \delta'(x{-}y)  
\nonumber \\ 
\{\jh_+^a(x),\jh_-^b(y)\} & =  & \half f^{abc}\left( 
\jh_+^c(x)  + \jh_-^c(x)  + \half i h^c_- (x) 
\right) \dxy \ .
\end{eqnarray}
We will also need the fermion equation of motion
\be
\del_- \psi_+ = {1 \over 2} [ \jh_- , \psi_+ ]
\quad \Rightarrow \quad
\del_+ \psi_+ = 2 \psi' + {1 \over 2} [ \jh_- , \psi_+ ]
\ee
to express all currents in terms of good canonical variables,
involving only {\em space\/} derivatives of the fermions.

\subsection{Poisson bracket algebra for currents built from 
symmetric traces}

Following the same route as for the bosonic PCM, we shall investigate 
currents built from symmetric-trace type invariants.
In principle all others can be expressed in terms of these (albeit in
a rather inconvenient, non-polynomial way for the case of the Pfaffian 
invariant).
After substituting in favour of the modified quantities 
just introduced, the odd-parity 
currents are:
\begin{eqnarray}
\F^-_{m+{1\over2}} & = & s_{a_1 \ldots a_m a_{m+1}} h_+^{a_1} \ldots 
h_+^{a_m} \psi_+^{a_{m+1}} = 
\tr (\psi_+^{2m+1})
\nonumber \\
\B^-_{m+1} & = & s_{a_1 \ldots a_m a_{m+1} }
h_+^{a_1} \ldots h_+^{a_m} \jh_+^{a_{m+1}} = 
\tr (\psi_+^{2m} \jh_+)\,.
\end{eqnarray}
The even-parity currents are considerably more complicated:
\begin{eqnarray}
\F^+_{m+{1\over2}} 
& = & s_{a_1 \ldots a_m a_{m+1}} h_+^{a_1} \ldots h_+^{a_m} \left(
\jh_+^{a_{m+1}} -\half i \psi_+^{a_{m+1}} \right) = 
\tr (\psi_+^{2m-1} \jh_+) - {i \over 2} \F_{m+{1\over2}}^-
\\
\B^+_{m+1} & = & 
s_{a_1 \ldots a_m a_{m+1}} \left(2i \psi_+^{a_1} \psi_+^{'a_2}
-ih_+^{a_1} \jh_-^{a_2} +m \jh_+^{a_1} \jh_+^{a_2} -(m{-}1)i h_+^{a_1}
\jh_+^{a_2} \right) h_+^{a_3} \ldots h_+^{a_{m+1}} \nonumber
\end{eqnarray}
\be
= 2i \tr( \psi_+^{2m-1} \psi'_+ ) - i \tr(\psi_+^{2m} \jh_- )
+ \sum_{r=0}^{m-1} \tr (\psi_+^{2r} \jh_+ \psi^{2m-2-2r}_+ \jh_+ )
- (m{-}1) i \B^-_{m+1}\,.
\ee
To calculate the PBs of these currents one can begin in the obvious way, 
by repeated application of the Leibnitz rules.
The resulting expressions can be simplified by the use of completeness 
conditions in the relevant Lie algebra. 
In particular, if $X$ is any element of $so(N)$ or $sp(N)$ then 
$X^m$ also belongs to the Lie algebra when $m$ is odd, and hence
$\tr( X^m t^c) \tr( Y t^c) = - \tr (X^m Y)$ for any $Y$.
The algebra $su(N)$ works a little differently, since the matrix must be
traceless in order to apply the completeness condition.
In this case we can write instead 
$\tr( X^m t^c) =
\tr( (X^m - { 1\over N} \tr (X^m) 1\, ) t^c )$
and then 
$\tr( X^m t^c) \tr( Y t^c) = - \tr(X^m Y) + {1\over N} 
\tr X^m \tr Y$ for any integer $m$.
To obtain the results given below it is also necessary to
take particular care with Fermi statistics and combinatoric factors,
and to make use of the cyclic properties of the trace to show that
certain terms vanish. 

The simplest brackets to calculate are those of the odd-parity currents 
amongst themselves, which can be shown to vanish: 
\begin{eqnarray}
\{ \F_{m+{1\over2}}^- (x) , \F_{n+{1\over2}}^- (y) \} & = & 0 
\nonumber \\
\{ \F_{m+{1\over2}}^- (x) , \B_{n+1}^- (y) \} & = & 0
\nonumber \\
\{ \B_{m+1}^- (x) , \B_{n+1}^- (y) \} & = & 0
\label{minuscurr}
\end{eqnarray}
For the odd-parity with even-parity currents we find: 
\bea
\{ \F_{m+{1\over2}}^- (x) , \F_{n+{1\over2}}^+ (y) \} &=& 
(2m{+}1) i \, \B^-_{m+n} \, \dxy \nonumber\\
\{ \B_{m+1}^- (x) , \F_{n+{1\over2}}^+ (y) \} &=&
- 2 \, \F^-_{m+n-{1\over2}} \, \dpxy -{ 2(2n{-}1) \over 2m{+}2n{-}1}
\, \F^{- \, \prime}_{m+n-{1\over2}} \, \dxy \nonumber\\
\{ \F_{m+{1\over2}}^- (x) , \B_{n+1}^+ (y) \} &=&
-2 (2m{+}1) \, \F^-_{m+n-{1\over2}}  \, \dpxy -{ 4n(2m{+}1) \over
2m{+}2n{-}1} \, \F^{- \, \prime}_{m+n-{1\over2}} \,  \dxy \nonumber \\
\{ \B_{m+1}^- (x) , \B_{n+1}^+ (y) \} &=& -4(m{+}n) \B^-_{m+n} \,
\dpxy -4n \, \B^{- \, \prime}_{m+n} \, \dxy
\label{mixedcurr}
\eea
where for clarity we have omitted the argument $x$ from all currents
appearing on the right-hand side.
The most difficult brackets to calculate are those of the 
even parity currents with  
themselves. After some effort we obtain the results:
\vfill\eject
\begin{eqnarray}
\{ \F_{m+{1\over2}}^+ (x) , \F_{n+{1\over2}}^+ (y) \} 
&=& i \B^+_{m+n} \dxy 
+{1 \over N} 
\left [-i\B_m^-\B_n^- \, + 2\F_{m-{1\over2}}^- \F^{-\,\prime}_{n-{1\over2}} \,
\right ]\dxy \nonumber \\
&& 
+ {2 \over N} \F_{m-{1\over2}}^- \F_{n-{1\over2}}^- \,  
\dpxy 
\nonumber\\
\{ \F_{m+{1\over2}}^+ (x) , \B_{n+1}^+ (y) \} &=&
-2(2m{+}2n{-}1) \F^+_{m+n-{1\over2}} \, \dpxy
-4n \F^{+ \, \prime}_{m+n-{1\over2}} \, \dxy 
\nonumber\\
&&+ {4n \over N} \
\left [ {1\over 2n{-}1} \B^-_m \F^{- \, \prime}_{n-{1\over2}} 
+ \F^-_{m-{1\over2}} \B^{- \, \prime}_n \right ] \, 
\delta (x{-}y).
\nonumber \\
&&+ {1 \over N} \left [ 2 \B^-_m \F^-_{n-{1\over2}} 
+ 4n \F^-_{m-{1\over2}} \B^-_n \, 
\right ] \, \delta' (x{-}y)
\nonumber\\
\{ \B_{m+1}^+ (x) , \B_{n+1}^+ (y) \} &=& -4(m{+}n) \B^+_{m+n} 
\dpxy -4n \B^{+ \, \prime}_{m+n} \dxy 
\nonumber \\
&& \mbox{\hskip -64pt} 
+\frac{1}{N} \left[ 8nm \B^-_m \B^{- \, \prime}_n 
-\frac{8ni}{(2n{-}1)(2m{-}1)} \F^{- \, \prime}_{m-\frac{1}{2}} \F^{- \,
\prime}_{n-\frac{1}{2}} +
\frac{8ni}{2n{-}1} \F^-_{m-\frac{1}{2}} \F^{- \, \prime
\prime}_{n-\frac{1}{2}} \right] \dxy 
\nonumber \\
&& \mbox{\hskip -36pt} 
+\frac{1}{N}
\left[
8nm \B^-_m \B^-_n 
-\frac{4i}{2m{-}1} \F^{- \, \prime}_{m-\frac{1}{2}}
\F^-_{n-\frac{1}{2}} 
+\frac{4i(4n{-}1)}{2n{-}1}
\F^-_{m-\frac{1}{2}} \F^{- \, \prime}_{n-\frac{1}{2}} \right] \dpxy
\nonumber \\
&& +\frac{4i}{N} \F^-_{m-\frac{1}{2}} \F^-_{n-\frac{1}{2}} 
\delta'' (x{-}y)
\label{pluscurr}
\end{eqnarray}
Once again, all fields on the right-hand side are at argument $x$.
The terms with $1/N$ coefficients occur only for the algebra 
$su(N)$, since for the other algebras $m$ and $n$ are always
odd.

Important consistency checks of these complicated calculations 
come from supersymmetry.
The current $\F^+_{3/2}$ is precisely the Noether current
for supersymmetry, and so
\[
\int dx \, \{ \F^+_{3/2} , X \} ,
\]
where $X$ is some current $\F^\pm_{n+1/2}$ or 
$\B^\pm_{n+1}$, must reproduce 
the transformations (\ref{susyhol}) applied to these quantities.
(In comparing the results one should 
remember that $\del_- \alpha = 0$ implies $\del_+ \alpha = 2 \alpha'$.) 
An even more powerful restriction arises if we take $X$ to be 
a Poisson bracket of currents.
Let us suppose (using an obvious notation) that we have calculated 
a PB of the form
$\{ \F_1 (x) , \F_2 (y) \}$. We can work out its variation under 
supersymmetry directly, but by the Leibnitz rule for PBs this is 
also proportional to the combination 
$\{ \B_1 (x) , \F_2 (y) \} - \{ \F_1 (x) , \B_2 (y) \}$, giving a 
non-trivial relationship between the latter two brackets.
Similarly, if we have calculated 
$\{ \B_1 (x) , \F_2(y) \}$, we can apply supersymmetry to relate
$\{ \B_1 (x) , \B_2(y) \}$ to $\{ \F'_1 (x) , \F_2 (y) \}$.
All the results above are consistent with such considerations.

\subsection{Commuting charges---by direct calculation}

We will now search for commuting sets of {\em bosonic\/} conserved charges,
beginning from 
\[ 
B_n^- = \int dx \, \B_{n+1}^- (x)
\ , \qquad
B_n^+ = \int dx \, \B_{n+1}^+ (x)
\]
based on the symmetric trace invariants used in the current algebra above.

There are a number of reasons why it seems natural {\em not\/} 
to consider fermionic charges in the same way.
For one thing, we explained in the last section that it is 
quite possible for a bosonic charge to survive quantisation without 
being accompanied by a superpartner.
Even if a fermionic charge were to survive quantisation along with its 
bosonic partner, it is not clear that it is very interesting to find 
`commuting' sets. This is because `commuting' for these classical 
charges really means `vanishing graded Poisson brackets', and if such an 
algebra is unmodified quantum-mechanically, the fermionic charges will obey
$F^2 = 0$. In a quantum theory with a positive-definite Hilbert space,
such charges can only be represented trivially.
One might then be prompted to consider other possibilities for the 
fermionic charge algebra, with $F^2 \neq 0$, and indeed such 
behaviour is evident already at the classical level in the PBs of the
currents $\F^+_{n+1/2}$. These are in some sense higher-spin
versions of supersymmetry, with $F^2 \sim B$ (schematically).
While such possibilities are certainly interesting, they lie 
beyond our immediate goals in this paper.

Let us turn then to the PB algebra of the bosonic charges
which is easily found to be 
\begin{eqnarray}
\{ B^-_m , B^-_n \} & = & 0 \nonumber\\
\{ B^-_m , B^+_n \} & = & 0 \nonumber\\
\{ B^+_m , B^+_n \} & = & {8mn \over N} \int dx \left ( 
\B_m^- \B_n^{- \, \prime} - 
{2i \over (2m{-}1)(2n{-}1)} \F^{- \, \prime}_{m-{1\over2}} 
\F^{- \, \prime}_{n-{1\over2}}  
\right ) \label{bosPBs}
\end{eqnarray}
The terms on the right-hand-side vanish for the algebras $so(N)$ and 
$sp(N)$, though not for $su(N)$.
This is highly reminiscent of the problems we were faced with in the 
bosonic theory. A further similarity is that 
our discussion so far is based on trace-type invariants,
and so omits the Pfaffian in $so(2 \ell)$.

Taking the same approach as before, 
we will first search for modifications of the even-parity currents 
which will yield commuting charges for $\Lg = su(N)$.
The simplest possibility is to add terms bilinear in
currents $\B^-_m$ and $ \F^-_{m-1/2}$, since these will naturally provide
contributions with the same structure as the unwanted terms 
we are hoping to cancel. In addition the trivial PBs of  
the negative parity currents implies that such modifications will not 
spoil the desirable property that the second bracket 
in (\ref{bosPBs}) vanishes.

Starting from a general ansatz, one finds after 
some algebra that the new currents
\be\label{curlyK}
\K^+_{m+1} = \B^+_{m+1}  -  {m \over N} \sum_{p=2}^{m-1} \B^-_p \B^-_{m-p+1}
-  {m \over N} \sum_{p=2}^{m-1} {2i \over (2p{-}1)(2m{-}2p{+}1) }
\F^-_{p-{1\over2}} \F^{- \, \prime}_{m-p+{1\over2}}
\qquad
\ee
have exactly the desired properties: {\it i.e.} the corresponding charges 
$K^+_n = \int dx \, \K_{n+1}^+$ obey 
\[
\{ B^-_m , B^-_n \} = \{ B^-_m , K^+_n \} = \{ K^+_m , K^+_n \} = 0 \ .
\]
These equations represent a highly over-determined system of conditions
for the coefficients of the new terms, so it is quite non-trivial 
that these have the correct properties.
In fact the result applies not just for $su(N)$,
but also for $so(N)$ and $sp(N)$, with the $1/N$ 
in the formulas above replaced by an arbitrary real number.
This further reinforces the analogy with the bosonic case.

It is natural to ask whether the additional terms 
correspond to anything simple in terms of the 
invariant tensors underlying the conserved currents.
The change from $\B^+_{m+1}$ to $\K^+_{m+1}$ 
actually amounts to a replacement 
\be\label{mod} 
\Lambda^{(2m)}_{a_1 \ldots a_{2m}} \rightarrow 
\Lambda^{(2m)}_{a_1 \ldots a_{2m}} 
- {1 \over N} \sum_{p=2}^{m-1} \Omega^{(2p-1)}_{[a_1 \ldots a_{2p-1}} 
\Omega^{(2m-2p+1)}_{a_{2p} \ldots a_{2m-1}] a_{2m}} 
\ee
which is easily checked for the fermionic terms in (\ref{curlyK}) (up to some 
irrelevant total derivatives) and which can be verified for 
the bosonic modifications too. 
Using detailed relationships between the 
invariant tensors which are derived in one of the appendices, section 9,
this is found to correspond to a replacement of 
the underlying symmetric tensor 
\[ s^{(m+1)}_{a_1 \ldots a_{m+1}} \rightarrow 
k^{(m+1)}_{a_1 \ldots a_{m+1}} \ ,
\]
precisely the set introduced in our treatment of the 
bosonic PCWZM in section 2. 

While this may seem very satisfactory, we must emphasise that 
the $k$-tensors were introduced in the bosonic theory to simplify the 
algebra of charges resulting from (\ref{JPBs}). A priori, there is no reason
to expect such a direct link with the considerably more 
complicated current algebra (\ref{pluscurr}) and it is therefore 
puzzling why the $k$-tensors should provide the 
required simplification in the SPCM too.
In the next section we shall resolve this puzzle by establishing a 
link with the earlier, bosonic, current algebra.
This approach also has the advantage of working for a general 
invariant tensor, so that the Pfaffian
charge in $so(2 \ell)$ can be treated in exactly the same way as the other
primitive invariants.

\subsection{Commuting charges---by comparison with bosonic PCM}

It is convenient to modify our notation very slightly.
We now take $k_{a_1 a_2 \ldots a_n a_{n+1}}$ to be any of 
the symmetric invariant tensors introduced in section 2 
via (\ref{kdef}), {\em or\/} the Pfaffian invariant in $so(2\ell)$ 
(previously written $p$ in (\ref{pfaff})).
This means that the $k$-tensors are now a complete
set of primitive invariants for any algebra.
Now denote the bosonic currents (\ref{oddbcomp}) and (\ref{evenbcomp})
with $d^{(n+1)} = k^{(n+1)}$, by $\K_n^-$ and $\K_n^+$ respectively.
The corresponding charges will be written
\[ 
K_n^- = \int dx \, \K_{n+1}^- (x) = 
\int dx k_{a_1 a_2 \ldots a_{n+1}} \jh_+^{a_1} h_+^{a_2} 
\ldots h_+^{a_{n+1}} 
\]
and 
\[
K_n^+ = \int dx \, \K_{n+1}^+ (x) = U_n +V_n +W_n- i (n{-}1) K^-_{n}
\]
where 
\bea
U_n &=& 2i\int dx \, k_{a_1 a_2 \ldots a_{n+1}} \psi^{a_1}_+
    {\psi^{a_2}_+}' h_+^{a_3} \ldots h_+^{a_{n+1}} \ , \nonumber \\
  V_n &=& n\int dx \, k_{a_1 a_2 \ldots a_{n+1}} \jh_+^{a_1} \jh_+^{a_2}
    h_+^{a_3} \ldots h_+^{a_{n+1}} \ , \nonumber \\
  W_n &=& -i \int dx \, k_{a_1 a_2 \ldots a_{n+1}} \jh_-^{a_1}
    h_+^{a_2} \ldots h_+^{a_{n+1}} \ . \nonumber 
\eea
(Since the $\Omega$ tensors are unique, $K_n^-$ is identical
to $B_n^-$ of the last subsection when the underlying $k$-tensor
is not the Pfaffian.) 
We will prove below that these charges commute:
\be\label{atlast}
\{ K^-_m , K^-_n \} = \{ K^-_m , K^+_n \} = \{ K^+_m , K^+_n \} = 0 \ .
\ee

One useful approach to these rather complicated calculations is to 
introduce the quantities 
\[ \h^a = h_+^a + \alpha \jh^a_+ \qquad {\rm and} \qquad
\tilde \h^a = h_+^a + \beta \jh^a_- 
\]
and to observe that
\begin{eqnarray}
K^-_n & = & {1 \over {n{+}1}} 
\int dx \, k_{a_1 a_2 \ldots a_{n+1}} \h^{a_1} \h^{a_2} \ldots 
\h^{a_{n+1}} \, \Big |_\alpha \label{KH}\\
V_n & = & {2 \over {n{+}1}} 
\int dx \, k_{a_1 a_2 \ldots a_{n+1}} \h^{a_1} \h^{a_2} \ldots 
\h^{a_{n+1}} \, \Big |_{\alpha^2} \label{VH} \\
W_n & = & - {i \over {n{+}1}} 
\int dx \, k_{a_1 a_2 \ldots a_{n+1}} \tilde \h^{a_1} \tilde \h^{a_2} \ldots 
\tilde \h^{a_{n+1}} \, \Big |_\beta \label{WH}
\end{eqnarray}
This provides a convenient way of handling the various combinatorial 
issues which arise. Furthermore, the 
PB algebra of the new currents has a structure similar to that encountered 
in the bosonic PCM: 
\begin{eqnarray} 
\{\h^a(x),\h^b(y)\} & = & f^{abc} \left ( ih^c_+ + \half\alpha^2( 3\jh_+^c
    - \jh_-^c) \right ) \delta(x{-}y) +2 \alpha^2 \delta^{ab}
    \delta'(x{-}y) \label{HPBs} \\
\{\h^a(x),\tilde \h^b(y)\} & = & f^{abc} \left ( ih^c_+ + \half\alpha \beta
( \jh_+^c + \jh_-^c)  + \quarter \alpha \beta i h^c_- \right ) \delta(x{-}y) 
\nonumber
\end{eqnarray}

The simplest computation is the bracket of two odd-parity charges,
$K^-_m$ and $K^-_n$, which is proportional to
\[
\left\{ \, \int dx \, 
k_{a_1a_2 \dots a_{m+1}}
    \h^{a_1} \h^{a_2} \dots \h^{a_{m+1}} , \int dy \, k_{b_1b_2
    \dots b_{n+1}} \h^{b_1} \h^{b_2} \dots \h^{b_{n+1}} \, \right\}
    \Big |_{\alpha^2} 
\]
The only surviving contribution is
\[
\int d x \, k^{(m+1)}_{a_1 a_2 \ldots a_m c} \, 
h^{a_1}_+ h^{a_2}_+ \ldots h^{a_m}_+ \, 
k^{(n+1)}_{b_1 \ldots b_{n-1} b_n c} \,
h^{b_1}_+ \ldots h^{b_{n-1}}_+ h^{b_n \, \prime}_+ 
\]
but this integrand vanishes due to invariance of the 
$k$-tensors, taken together with the fact that
$h_+^a = \half f^{abc} \psi_+^b \psi_+^c$. Indeed, 
if the expression is re-written in terms of $\Omega$ tensors, 
its vanishing is equivalent to the identity
\[
\Omega^{(2m+1)}_{ c [a_1 \ldots a_{2m}} \,
\Omega^{(2n+1)}_{ b_1 \ldots b_{2n-1} ] b_{2n} c} = 0 \ .
\]
(This can be proved by writing the $\Omega$-tensors as in (9.8) and
using the invariance condition (9.2).)
Thus $\{ K^-_m , K^-_n \} = 0$, as claimed.

Turning next to the bracket of $K^-_m$ with $K^+_n$, it
suffices to consider the brackets of the odd-parity charge with 
$U_n$, $V_n$ and $W_n$. For the first of these, we need a 
lemma:
\[ \{ U_n , h^a_+ (x) \} =  \{ U_n , \h^a (x) \} = - 4( k_{a_1 \ldots a_n a}
h^{a_1}_+ \ldots h^{a_n}_+ )'(x)  \ .
\]
Then from (\ref{KH}) we find
\[ \{ K^-_m , U_n \} = \int d x \, 4mn \, 
k^{(m+1)}_{a_1 \ldots a_{m-1} a_m c} \,
h^{a_1}_+ \ldots h^{a_{m-1}}_+ \jh^{a_m}_+ \,
k^{(n+1)}_{b_1 \ldots b_{n-1} b_n c} \,
h^{b_1}_+ \ldots h^{b_{n-1}}_+ h^{b_n \, \prime}_+ \ .
\]
The remaining brackets we need are
\begin{eqnarray*}
\{ K^-_m , V_n \} & = & - \int d x \, 4 mn \,  
k^{(m+1)}_{a_1 \ldots a_{m-1} a_m c} \,
h^{a_1}_+ \ldots h^{a_{m-1}}_+ h^{a_{m} \, \prime}_+ \, 
k^{(n+1)}_{b_1 \ldots b_{n-1} b_n c} \, 
h^{b_1}_+ \ldots h^{b_{n-1}}_+ \jh^{b_n}_+
\\
&& + \int dx \, n \,
k^{(m+1)}_{a_1 \ldots a_m a} \, 
h^{a_1}_+ \ldots h^{a_{m}}_+ \,
k^{(n+1)}_{b_1 \ldots b_{n-1} b_n b} \,
h^{b_1}_+ \ldots h^{b_{n-1}}_+ \jh^{b_n}_+ f^{abc} \jh^c_- 
\end{eqnarray*}
which follows from (\ref{KH}) and (\ref{VH}),
and
\[
\{ K^-_m , W_n \} = - \int d x \, mn \, k^{(m+1)}_{a_1 \ldots a_{m-1} a_m a}
h^{a_1}_+ \ldots h^{a_{m-1}}_+ \jh^{a_m}_+ \,
k^{(n+1)}_{b_1 \ldots b_{n-1} b_n b} \, 
h^{b_1}_+ \ldots h^{b_{n-1}}_+ \jh^{b_n}_- f^{abc} h_+^c
\]
which follows from (\ref{KH}) and (\ref{WH}).
In each of these calculations it is necessary to make extensive 
use of the invariance conditions for the $k$-tensors. These same conditions
then imply that the total contribution is
\[ \{ K^-_m , K^+_n \} = \{ K^-_m , U_n \} + \{ K^-_m , V_n \} 
+ \{ K^-_m , W_n \} = 0 \ .
\]

Finally we come to the lengthiest calculation: the bracket of two even-parity 
charges. Given the properties of the odd-parity charges, it suffices 
to consider the brackets of the quantities $U$, $V$ and $W$ amongst
themselves, presenting us with six different expressions to evaluate.
Some of these are very similar to the calculations we have already
sketched above. In particular, we find
$\{ W_m , W_n \} = \{ U_m , W_n \} + \{ W_m , U_n \} = 0$.
The non-trivial contributions can then be usefully divided into two:
the terms 
\be
\{V_m,V_n\} + \{U_m,V_n\} +\{V_m,U_n\} + 
\{W_m,V_n\} +\{V_m,W_n\} 
\label{first_terms}
\ee
which can be treated using the formulas 
(\ref{VH}) and (\ref{WH}) together with the lemma introduced earlier;
and a single remaining term $\{U_m,U_n\}$ which must be evaluated by
other means.
We will now show that both sets of terms vanish, by comparing with
the known results for the bosonic models. 

After some work it can be shown that (\ref{first_terms}) 
is equal to 
\[
\frac{4}{(m{+}1)(n{+}1)} \left\{ \, \int dx \, 
k_{a_1a_2 \dots a_{m+1}}
    \h^{a_1} \h^{a_2} \dots \h^{a_{m+1}} , \int dy \, k_{b_1b_2
    \dots b_{n+1}} \h^{b_1} \h^{b_2} \dots \h^{b_{n+1}} \, \right\}
    \Big |_{\alpha^4} 
\]
This expression obviously contains one of the desired brackets,
$\{ V_m , V_n \}$, but it also generates other terms which turn 
out to match exactly the remaining contributions in (\ref{first_terms}).
Now we need only compare the current algebra (\ref{HPBs}) of the 
$\h^a$ with (\ref{LCPBs}) in the bosonic PCM to understand why this
expression vanishes. Exactly the same arguments (as given in section 2 and 
\cite{EHMM2}) ensure that the ultralocal terms will not contribute,
and so we have the same charge algebra as for 
the bosonic PCM, up to an overall constant arising from the 
coefficients of the $\delta'$ terms. 
The tensors $k$ were chosen precisely to ensure that the charge PBs vanished 
in the bosonic PCM. Hence (\ref{first_terms}) also vanishes.

To complete the computation of the even-parity charge brackets it remains 
to consider 
\[
\{ U_m , U_n \} =   
-16mni \int dx \, k^{(m+1)}_{a_1 \ldots a_{m-1} bc} k^{(n+1)}_{a_m \ldots
    a_{m+n-2}dc} h_+^{a_1} \ldots h_+^{a_{m+n-2}} \psi_+^{'b}
    \psi_+^{'d} \, , 
\]
which is once again arrived at by extensive use of invariance conditions.
The antisymmetry in $b$ and $d$ imposed by $\psi_+^{'b} \psi_+^{'d}$
allows us to write this, up to a factor, as
\[
\int dx \, {k^{(m+1)}_{(a_1 \ldots a_{m-1}b}}^c
    k^{(n+1)}_{a_m \ldots a_{m+n-2})dc} h_+^{a_1} \ldots h_+^{a_{m+n-2}}
    \psi_+^{'b} \psi_+^{'d} \,.
\]
But now recall that the vital property of the $k$-tensors 
that guarantees commuting charges in
the bosonic PCM can be expressed as (\ref{keqn}).
This immediately implies that the bracket $\{ U_m , U_n \}$ vanishes.

We have now established (\ref{atlast}).
Notice that most of the arguments---including all those underlying the
vanishing of the PBs of the odd-parity charges---did 
not involve any special property of the $k$-tensors (beyond 
their invariance). The special nature of the $k$ tensors was used  
at precisely two points above in showing that the even-parity charges
have vanishing PBs too. In making comparisons with the 
current algebra of the bosonic PCM, we have clarified 
why the same tensors arise in the SPCM.

\section{Summary and conclusions}

In the bosonic PCWZM, there are infinitely many holomorphic 
conservation laws (\ref{gencons}) based on symmetric invariant tensors.
{}From amongst this set, it is possible to define commuting local charges
based on the particular symmetric tensors $k$ defined in section 2.
There is an infinite sequence for each primitive invariant, with spins
repeating modulo the Coxeter number of the algebra.
All this is completely independent of the coefficient of the WZ term.

In the supersymmetric versions of these models\footnote{We discussed 
mainly the SPCM, but the extension to the supersymmetric PCWZM 
should be obvious in view of our detailed treatment of the bosonic theories.}, 
there are finitely many independent holomorphic conservation laws 
(\ref{oddgen},\ref{evengen}). As explained in section 3, 
they are based on antisymmetric invariant tensors, which can nevertheless 
be related to symmetric {\em primitive\/} invariants in the algebra. 
This leads to bosonic conserved charges with spins exactly equal to the 
exponents, but with no repetition modulo the Coxeter number. 
These charges commute with one another when the symmetric invariants 
are chosen to be precisely the same tensors $k$ that arose in the 
bosonic theory.  

There is no direct relationship between the currents (\ref{gencons})
and (\ref{oddgen},\ref{evengen}) and in fact the latter vanish 
when the fermions are set to zero. A simple and direct relationship exists 
only between much larger sets of holomorphic currents which are
special features of the WZW and super WZW models. 
A rather subtle indirect relationship can be established between
the underlying current algebras (\ref{JPBs}) and 
(\ref{minuscurr})-(\ref{pluscurr}), however,
which explains the importance of the same set of tensors $k$ for both the
bosonic and supersymmetric theories.

{\bf Acknowledgments}

We thank Jose Azc\'arraga, Patrick Dorey and G\'erard Watts for discussions.
The research of JME is supported by a PPARC Advanced Fellowship, and by
NSF grant PHY98-02484.
NJM thanks Pembroke College Cambridge for a Stokes Fellowship, during
which early stages of this work were carried out. 
MH is grateful to St. John's College, Cambridge for a Studentship.
AJM thanks the 1851 Royal Commission for a Research Fellowship.

\vspace{0.2in}

\section{Appendix: Poisson brackets in the PCWZM and SPCM}

A general $\sigma$-model with WZ term can be described by a lagrangian
of the form
\bea
{\cal L} = {1 \over 2} \, g_{ij} (\phi) \, \del_\mu \phi^i \, \del^\mu \phi^j
+ {1 \over 2} \, b_{ij} (\phi) \, \varepsilon^{\mu \nu} \, 
\del_\mu \phi^i \, \del_\nu \phi^j \\
= {1 \over 2} \, g_{ij} (\phi) \, ( \, \dot \phi^i \, \dot \phi^j
- \phi'^{\, i} \, \phi'^{\, j} \, )
+ \, b_{ij} (\phi) \, \dot \phi^i \, \phi'^{\, j}
\eea 
where $\phi^i$ are coordinates on some target manifold equipped with a 
metric $g_{ij} (\phi)$ and antisymmetric tensor field $b_{ij} (\phi)$. 
The momenta conjugate to the fields $\phi^i$ are
\[
\pi_i = 
\hat \pi_i + b_{ij} \phi'^{\, j} 
\qquad {\rm where} \qquad \hat \pi_i = g_{ij} \dot \phi^j
\]
by definition.
These obey the standard non-vanishing equal-time PBs
\[
\{ \, \phi^i (x) , \, \pi_j (y) \, \} = \delta^i{}_j \, \delta (x{-}y)
\, .
\]
A short calculation reveals that 
\[
\{ \hat \pi_i (x) , \hat \pi_j (y) \} = h_{ijk} \, \phi'^{\, k} \, 
\dxy
\qquad {\rm where} \qquad h_{ijk} = \del_i b_{jk} + \del_j b_{ki} + 
\del_k b_{ij}
\]

Now consider a (non-conserved) current 
\[
E^a_\mu = E^a_i \, \del_\mu \phi^i
\]
where $E^a_i (\phi)$ are vielbeins on the target manifold satisfying
\[
E^a_i E^a_j = g_{ij}
\]
In terms of the canonical coordinates $\phi^i$ and $\pi_i$ we have
\[
E^a_0 = E^a_i g^{ij} \hat \pi_j \, , \qquad E^a_1 = E^a_i \phi'{}^{\, i} \, . 
\]
The PB algebra of these currents can now be calculated routinely,
although the general result requires some effort and is not 
particularly illuminating.

Important simplification occurs for the special case of a group
manifold, with currents defined by the (right-transforming) vielbeins
\[ 
E^{a}_i = \tr(t^a g^{-1} \del_i g) \, ,
\qquad 
{\rm obeying } 
\qquad
\del_{[i} E_{j]} = E_{[i}E_{j]} \, .
\]
Let us also choose the WZ term to be related to the structure constants 
\[
h^{ijk} E^{a}_i E^b_j E^c_k = - \lambda f^{abc}
\]
with $\lambda$ some constant. This is precisely the PCWZM  
defined, in different notation, in the main text.
For this case, the results of the current algebra calculations simplify
to give
\begin{eqnarray*}
\{ E^a_0(x),E^b_0(y) \} & = & f^{abc} \, 
(\, E^c_0  - \lambda E^c_1 \, ) \, \dxy \\
\{ E^a_0(x),E^b_1(y) \} & = & f^{abc} E^c_1 \, \delta (x{-}y) + \delta^{ab} 
\dpxy \\
\{ E^a_1(x),E^b_1(y) \} & = & 0
\end{eqnarray*}
Notice that the only effect of the WZ term is the contribution
proportional to the constant $\lambda$.
Translating to light-cone components gives
\begin{eqnarray*}
\{ E^a_\pm(x),E^b_\pm(y) \} & = & 
{1 \over 2}
f^{abc} \, ( \, (3{\mp}\lambda) E^c_\pm  - (1{\mp}\lambda) E^c_\mp \, )
\, \dxy 
\, \pm \, 2 \dpxy\\
\{ E^a_+(x),E^b_-(y) \} & = & {1\over2}f^{abc} \, ( \,
(1{-}\lambda) E^c_+ + (1{+}\lambda) E^c_- \, ) \, \delta (x{-}y) 
\end{eqnarray*}

In the PCWZM the conserved current has components
$j_\pm = j_0\pm j_1 = (1\pm\lambda)(E_0\pm E_1)=(1\pm\lambda)E_\pm$.
The expressions given in the main text, with $\kappa =1$,
now follow immediately. 

Now we consider the supersymmetric PCM, using the same notation for
coordinates $\phi^i$ and vielbeins $E^a_i$ on the group as above, 
but with fermions $\psi_\pm^a$ valued in the Lie-algebra 
({\it i.e.\/ } carrying tangent-space indices) as in the main text. 
To determine the Poisson brackets, only the terms in the lagrangian 
involving time derivatives of the fields are important.
After re-writing the couplings 
${\rm Tr}( \psi_\pm [ g^{-1} \del_\mp g , \psi_\pm])$ in coordinate notation,
the only relevant terms are
\[
{1 \over 2} \, g_{ij} (\phi) \, \dot \phi^i \, \dot \phi^j
+ { i \over 2} \psi_+ \dot \psi_+ 
+ { i \over 2} \psi_- \dot \psi_- 
+ { i \over 2} E^a_j \dot \phi^j (h_+^a + h_-^a)
\] 
The brackets amongst the fermions are just those of a free theory,
with the standard normalizations for real fermions.
Moreover, they have vanishing brackets with the fields $\phi^i$ 
and with their conjugate momenta
\[
\pi_i = \hat \pi_i + { i\over 2} E^a_i (h^a_+ + h^a_-)
\qquad {\rm where} \qquad
\hat \pi_i = g_{ij} \dot \phi^j 
\]

The conserved currents in the SPCM have spacetime components
\begin{eqnarray} 
&&j^a_0 = E^a_i \dot \phi^i - {i \over 2} (h^a_+ + h^a_-)
      = E^{a \, i} \pi_i - i (h^a_+ + h^a_-)
\\
&&j^a_1 = E^a_i \phi'^{\, i} - {i \over 2} (h^a_+ - h^a_-)
\end{eqnarray}
It is now straightforward to calculate the algebra 
by comparing with the result for the bosonic
PCM $(\lambda=0)$ above, and using the results (\ref{fermpbs}) for the 
fermions. One finds (\ref{spcmpbs}) together with
\[
\{ j^a_0 (x) , \psi^b_\pm (y) \} = f^{abc} \psi_\pm^c \delta(x{-}y)
\, , \qquad
\{ j^a_1 (x) , \psi^b_\pm (y) \} = 
\pm {1 \over 2} f^{abc} \psi_\pm^c \delta(x{-}y) \ .
\]
Changing to light-cone components gives the expressions quoted in the 
text.

By combining the approaches above, the Poisson brackets of the super PCWZM
can be calculated in a similar fashion.

\section{Appendix: Conservation laws in superspace}

The superspace conservation equation (\ref{nonhol}) 
has component content (\ref{e1})-(\ref{e4}).
Only the first of these equations represents a 
conservation law, in general.
For the special case of a holomorphic current, however, there is an additional 
conserved quantity as in (\ref{chargepair}) and the pair are related by
supersymmetry (\ref{part}).
The first issue we wish to clarify here is how such 
a superpartner can arise in more general circumstances, 
including necessary and sufficient conditions for this to happen.

In order for  
(\ref{e2}) to give an additional conservation law
we require that $\beta_- = - \del_+ \omega_{---}$ for some (spin-3/2) 
fermion $\omega_{---}$, so that 
\be 
\del_- \alpha_+ + \del_+ \omega_{---} = 0 
\quad \Rightarrow \quad 
F^+ = \int (d x^+  \alpha_+ + d x^- \omega_{---} )
\ee
is a new conserved charge, which is 
related to $B$ in (\ref{scharge}) by supersymmetry.
But by applying a supersymmetry transformation, we find that 
the constraint $\beta_- = - \del_+ \omega_{---}$ is consistent 
only if $v = - \del_+ k_-$, for some (spin-1) boson $k_-$. 
Taken together, these imply 
\be\label{improve} 
J_- = - i D_+ K_- \, , 
\qquad {\rm where} \qquad 
K_- = k_- + i \theta^+ \alpha_- + i \theta^- \omega_{---} +
i \theta^+ \theta^- j_- \, . 
\ee
Thus a necessary and sufficient condition 
for the existence of a conserved charge $F^+$ whose variation 
under $Q_+$ gives $B$, is that we can write $J_- = - iD_+ K_-$ for some 
superfield $K_-$. The holomorphic case corresponds to the simplest  
possibility $K_{-} = 0$. 
There is also the independent possibility that we can construct a  
superpartner charge $F^-$ related to $B$ by $Q_-$, which arises if and 
only if $J_+ = -i D_- K_+$ for some superfield $K_+$.

Notice that when (\ref{improve}) is satisfied, we can 
re-express (\ref{nonhol}) in the form 
\be\label{nonsf}
D_- ( i \theta^+ J_+ ) - D_+ ( i \theta^+ J_- - K_- ) = 0 \, . 
\ee
This is also a superspace conservation equation, but the 
current components are not superfields.
We can construct a conserved quantity 
from this new equation by using the standard formulas in (\ref{scharge}),
and the result is $F^+$.
Our previous arguments ensuring invariance under supersymmetry 
of the conserved charge $B$ do not apply to $F^+$, 
because the current components in (\ref{nonsf}) 
involve $\theta$ explicitly. 

It is helpful to compare this with symmetries in ordinary (non-super) 
spacetime.
Any charge constructed entirely from fields, such as a momentum
generator or an internal symmetry generator, must commute with translations.
But charges which involve explicit dependence on $x$-coordinates, 
such as Lorentz generators, will not commute with translations.
Similarly, in superspace, any charge constructed entirely from
superfields will necessarily commute with supersymmetry.
But charges involving explicit $\theta$-dependence will not.

The second issue we would like to elaborate on is 
how this discussion applies to 
energy-momentum and supersymmetry.
As a consequence of translation invariance, Noether's Theorem 
guarantees the existence of a superfield conservation law 
of the general form (\ref{nonhol}) 
with the bosonic charge $B$ being energy-momentum
along some particular direction. 
For a supersymmetric theory we know there is 
a conserved superpartner $F$, namely a supersymmetry generator.
But to establish this we must also apply Noether's theorem to 
supersymmetry transformations.
Once this is done, we find that the 
definition of the translation superfield current can indeed 
always be improved so as to fulfill the condition (\ref{improve}),
in accordance with our general results.

The necessity of carrying out such an improvement in the conformal
case was discussed in \cite{clark81}, but in language pre-dating the modern 
development of conformal field theory.
In contemporary terminology, this is simply the statement that in a
superconformal field theory we can always improve the canonical 
super-energy-momentum tensor so that its
conservation becomes a holomorphic conservation equation.
To complete our discussion we will show how this improvement 
works in a general supersymmetric theory, whether conformal or not.

In the main text we deliberately avoided the raising and lowering of spinor 
indices. Now it will be more helpful to allow this possibility.
We shall distinguish vector indices 
$\mu, \nu , \ldots $ and spinor indices $a, b , \ldots$ with
the summation convention applied to all contracted upper and lower indices.
The rule for raising and lowering spinor indices is 
$F^\pm = \pm F_\mp$. Thus, for example, the standard 
superspace current conservation equation reads
\[
D_a J^a = D_+ J^+ + D_- J^- = D_+ J_- - D_- J_+ = 0 \ .
\]

Consider a field theory in superspace, described by a
superfield lagrangian
${\mathcal L}(\Phi,D_a\Phi)$.
Under the action of 
graded transformations which change the superfield by
$ \delta\Phi(x^\mu ,\theta^a)$, one finds that, on
using the equations of motion, the variation of the lagrangian can be
expressed in the form 
\[
D_a(\frac{\partial\mathcal{L}}{\partial D_a\Phi} 
\delta\Phi)-\delta{\mathcal{L}}=0 .
\]
Now the condition for invariance of the action is 
\[
\delta {\cal L} = D_a X^a \qquad \Rightarrow \qquad 
D_a J^a = 0 \, , \qquad {\rm with} 
\qquad J^a = \frac{\partial\mathcal{L}}{\partial
D_a\Phi} \delta \Phi - X^a \, ,
\]
where the first equation defines $X^a$.
This is the superspace form of Noether's Theorem.

Applying this to $x$-translations in the direction labelled $\mu$, 
gives 
\[
D_aT^a{}_{\mu}=0
\]
with $T^a{}_\mu$ a vector-spinor superfield. In detail: 
\be\label{tnoet}
D_a(\frac{\partial{\cal L}}{\partial D_a\Phi} 
\del_\pm \Phi)- D_\pm ( i D_\pm {\cal L}) =0
\ee
where we have made 
use of the superspace algebra $D_\pm^2 = -i \del_\pm$.
Applying Noether's Theorem to a supersymmetry labelled by a spinor
index $b$, we find 
\[
D_a S^a{}_b = 0
\]
where $S^a{}_b$ is a spinor-spinor superfield. By making repeated 
use of the fact that
$Q_\pm = D_\pm + 2i \theta^\pm \del_\pm$
we can write the current 
\be\label{snoet}
S^a{}_\pm = K^a{}_\pm - 2i \theta^\pm T^a{}_\pm \, , \qquad {\rm where} \qquad
K^a{}_b = \frac{\del {\cal L}}{\del (D_a \Phi)} D_b \Phi  + 
\delta^a_b {\cal L}
\ee

Comparing (\ref{snoet}) and (\ref{tnoet}) we see that 
\be\label{imp} 
T^\pm{}_\pm = - {i \over 2}D_a K^a{}_\pm \ 
\ee 
and it is this which allows us to improve the energy-momentum 
tensor in the way that we desire.
For example, considering 
translations or supersymmetries given by $ \mu = b = +$, 
it follows from (\ref{imp}) that we can define
an improved superfield current:
\be 
{\tilde T}^\pm{}_+ = T^\pm{}_+  + \frac{i}{2} D_\mp K^-{}_+
\ee
which satisfies
\be 
D_a {\tilde T}^a{}_+ = 0 \, , \qquad {\tilde T}^+{}_+ = - \frac{i}{2} D_+
K^+{}_+
\ee 
The last equation exactly meets the criterion (\ref{improve}).  
Similarly, for translations and supersymmetries given by
$\mu = b = -$ we have the independent improvement
\be 
{\tilde T}^\pm{}_- = T^\pm{}_-  + \frac{i}{2} D_\mp K^+{}_-
\ee
which satisfies
\be
D_a {\tilde T}^a{}_- = 0 \, , \qquad {\tilde T}^-{}_- = - \frac{i}{2} D_-
K^-{}_- \, .
\ee
The superpartner charges which arise are of course 
exactly the supersymmetry generators.

\section{Appendix : Commutation of local with non-local charges}

In this appendix we give some details of the vanishing of the
Poisson brackets of the local with the non-local charges for the
SPCM ($\lambda = 0$). Recall that the odd-parity charges
$F^-_{m-{1\over2}}$ and $B^-_m$ are integrals of the currents
\[
\Omega_{a_1 a_2 \ldots a_{2m+1} } \psi^{a_1}_+ \psi^{a_2}_+ \ldots
\psi^{a_{2m+1}}_+ \; , \qquad \Omega_{a_1 \ldots a_{2m} a_{2m+1} }
\psi^{a_1}_+ \ldots \psi^{a_{2m}}_+ j^{a_{2m+1}}_+ \,,
\]
respectively. Once we have proved that the non-local charges
commute with these, we may prove commutation for the even-parity
local charges $F^+_{m-{1\over2}}$ and $B^+_m$ by considering the
densities
\[
d_{a_1 a_2 a_3 \ldots a_{m+1}} j_+^{a_1} \psi_+^{a_2} h_+^{a_3}
\ldots h_+^{a_{m+1}} , \qquad d_{a_1 a_2 a_3 \ldots a_{m+1}} ( m
j_+^{a_1} j^{a_2}_+ + i \psi^{a_1}_+ \del_+ \psi^{a_2}_+ )
h_+^{a_3} \ldots h_+^{a_{m+1}} \,,
\]
(since they differ by terms proportional to the odd-parity
charges). Recalling the definitions $h_\pm = \psi_\pm^2$ given in
the text, we shall also write $h_0 = {1 \over 2} (h_+ + h_-)$ and
$h_1 = {1 \over 2} (h_+ - h_- )$. It is useful to introduce
quantities $b_\mu$ by writing the conserved current components in
the SPCM
\[
j_0=b_0- 2 i h_0 \;\;, \qquad j_1= b_1 - i  h_1 \,,
\]
The quantities $b_\mu$ satisfy the same Poisson brackets as the
currents in the bosonic PCM ((\ref{STPBs}, \ref{LCPBs}) with
$\lambda = 0$), while the relations (\ref{fermpbs}) can also
conveniently be re-written
\[
\{ h_\mu^a(x), h_\nu^b(y) \} = \half i f^{abc}
h_{|\mu-\nu|}^c(x)\delta(x-y)\,.
\]
for $\mu , \nu = 0, 1$.

Since all the local charges we have constructed are singlets in
the Lie algebra, the charge $Q^{(0)\,a}$ must commute with them;
this is also relatively simple to check directly. It therefore
remains to calculate the brackets with the first non-local charge,
which can be written
\[
Q^{(1)a} = \int \left\{ b_1^a(x) - 2 i h_1^a(x) - \half f^{abc}
j_0^b(x) \int^x j_0^c(y) \,dy\right\} dx  \ .
\]
In the calculations which follow, we shall use square brackets to
indicate the contributions arising from each of the three terms in
the formula for $Q^{(1)a}$ above.

The simplest bracket
\[
\{F_{m-{1\over2}}^-, Q^{(1)a} \} = 0
\]
can be calculated quite straightforwardly: the first term is
trivially zero, and the other two vanish by invariance of 
$\Omega$.

Next we consider the even-parity fermionic charge, constructed
using a $\Lambda$ tensor which is antisymmetric on all but one of
its $2m$ indices. 
\beaa \{F_{m-{1\over2}}^+, Q^{(1)c} \} & = & \left\{ \,\int
\Lambda_{a_1 a_2 \ldots a_{2m-1} b} \,\psi_+^{a_1} \psi_+^{a_2}
\ldots \psi_+^{a_{2m-1}} j_+^b \, dx\,, Q^{(1)c} \,\right\} \\ & =
& \int  dx\,\Lambda_{a_1 a_2 \ldots a_{2m-1} b} \left\{
( \,f^{a_1cd}\psi_+^d \psi_+^{a_2} \ldots \psi_+^{a_{2m-1}} 
+ \ldots + 
f^{a_{m-1}cd}\psi_+^{a_1} \ldots \psi_+^{a_{2m-2}} \psi_+^d \, ) 
j_+^b \right.\\ &&
\left.\;\;+ f^{bcd}\psi_+^{a_1} \psi_+^{a_2} \ldots
\psi_+^{a_{2m-1}} \left( \left[b_1^d\right]+\left[ -2 i h_1^d -
ih_0^d\right] +\left[b_0^d -2 i h_0^d\right] \right) \right\}\\ &
\propto & \int dx\, \Lambda_{a_1 a_2 \ldots a_{2m-1} b} f^{bcd}
\psi_+^{a_1} \psi_+^{a_2} \ldots \psi_+^{a_{2m-1}} h_+^d \\
& \propto & \int  dx\,f^{bcd} f^{da_{2m}a_{2m+1}}
\Lambda_{a_1 a_2 \ldots a_{2m-1}b} \psi_+^{a_1} \psi_+^{a_2}
\ldots \psi_+^{a_{2m+1}} \, .
\eeaa 
But, by the Jacobi identity and invariance:
\[
f^{bcd} f^{d}{}_{[e a} \Lambda_{a_1 a_2 ... a_{2m-1}]}{}^{b} = 
2f^{cd}{}_{[e} f^{db}{}_{a} \Lambda_{a_1...a_{2m-1}]}{}^{b} = 
-2 f^{cd}{}_{[e} \Omega^{d}{}_{a a_1...a_{2m-1}]} = 0 
\ .
\]

Now we move on to the rather tougher bosonic charges. (One could
deduce that these commute with $Q^{(1)a}$ by applying
supersymmetry to the fermionic results; we shall calculate them
directly.) 
To simplify the results it is wise to introduce 
some notation in advance. Since invariance of the $d$ tensors 
plays such an important role, we introduce the short-hand
\[
\underline{A}BC \ldots D \equiv  
f^{ba_1c}d_{a_1a_2a_3...a_{m+1}} A^c B^{a_2} C^{a_3} \ldots 
D^{a_{m+1}}
\,,
\]
Then, by symmetry of $d$, $\underline{A}BC \ldots D  =
B\underline{A}C \ldots D =BC\underline{A} \ldots D$ {\it etc}. 
The fact that $d$ is invariant may now be expressed:
\[
{\bf d}(ABC \ldots D)\equiv \underline{A}BC \ldots D + A\underline{B}C
\ldots D \, + \, \ldots \, + \, ABC \ldots \underline{D} = 0\, .
\]

Consider now the odd-parity bosonic charge:
\beaa 
\{Q^{(1)b}, B_m^-\} & = & \int dx\, d_{a_1a_2 \ldots a_{m+1}} 
\left\{ \, \,
\left[\, f^{ba_1c} b_1^c h_+^{a_2} \ldots h_+^{a_{m+1}} \, \right]\right.\\ 
&& \qquad + \,\left[\, -if_{ba_1c}(2h_1^c+ h_0^c) 
h_+^{a_2} \ldots h_+^{a_{m+1}} \right .\\ 
&&\qquad \qquad \qquad \left .
+ \, \, j_+^{a_1} (f^{ba_2c} h_+^c h_+^{a_3} \ldots h_+^{a_{m+1}}
+ \ldots + f^{ba_{m+1}c} h_+^{a_2} \ldots h_+^{a_m} h_+^c) \, \right]\\
&&\qquad + \, \, \left.\left[\, f^{ba_1c}(b_0^c - 2i h_0^c) h_+^{a_2}
\ldots h_+^{a_{m+1}} \, \right] \, \, \right\}
\eeaa
where we have made repeated use of invariance of $d$ to 
eliminate certain ultralocal and non-ultralocal terms.
The surviving terms written above can now be grouped together into two 
sets proportional to 
$ {\bf d}(h_+^{m+1})=0$ and ${\bf d}(j_+ h_+^m) = 0$.
Hence the bracket vanishes.

Finally we consider the bracket of $Q^{(1)a}$ with
\[
\int dx\,d_{a_1 a_2 a_3 \ldots a_{m+1}} \left( m j_+^{a_1}
j^{a_2}_+ + 2i \psi^{a_1}_+ \del_1 \psi^{a_2}_+  - i (j_-^{a_1} +
\half i h_-^{a_1}) h_+^{a_2}\right) h_+^{a_3} \ldots h_+^{a_{m+1}}
\, ,
\]
which we know differs from $B^+_m$ by a term proportional to
$B^-_m$. The resulting expression has three groups of three
$[\dots]$ terms, one for each of $j^2, \psi\del_1\psi$ and $(j_- +
\half i h_-) h_+$: 
\beaa 
\{Q^{(1)b}, B_m^+\} 
& = &\int dx\, d_{a_1a_2 \ldots a_{m+1}} 
\left\{ \, \, 
m\left[ \, ( \, f^{ba_1c}j_+^{a_2}+f^{ba_2c}j_+^{a_1} \, ) b_1^c 
h_+^{a_3} \ldots h_+^{a_{m+1}} \, \right] \right.\\ 
&&\;\;+ \, 
m\left[ \, -i (\, f^{ba_1c}j_+^{a_2} + f^{ba_2c}j_+^{a_1}\, )
(2h_1^c + h_0^c) h_+^{a_3}\ldots h_+^{a_{m+1}}\right.\\ 
&&\left.\;\;\qquad\qquad+ \, 
j_+^{a_1} j_+^{a_2} 
(\, f^{ba_3c} h_+^c h_+^{a_4} \ldots h_+^{a_{m+1}} \, + \ldots + \, 
f^{ba_{m+1}c} h_+^{a_3} \ldots h_+^{a_{m}} h_+^c \, ) 
\, \right]
\\ 
&&\;\; + \, 
m\left[ \, ( \, f^{ba_1c}j_+^{a_2} + f^{ba_2c}j_+^{a_1} \, ) \, j_0^c
h_+^{a_3} \ldots h_+^{a_{m+1}}\right] \\
&&\;\; + \; \; [0]+[0] + \left[-4if^{ba_1c}j_0^c h_+^{a_2} \ldots
h_+^{a_{m+1}}\right]\\ 
&&\;\;+ \, \left[ -if^{ba_1c} b_1^c h_+^{a_2}\ldots h_+^{a_{m+1}}\right] \\ 
&&\;\;+\,\left[
\, - \, i (b_0^{a_1}\!-\!b_1^{a_1}\!-\!2ih_0^{a_1})
( \, f^{ba_2c}h_+^c h_+^{a_3} \ldots h_+^{a_{m+1}} \, + \ldots + \, 
f^{ba_{m+1}c} h_+^{a_2} \ldots h_+^{a_m} h_+^c
\, )
\right .
\\
&& \qquad \qquad \qquad \qquad \qquad \qquad \qquad \qquad 
\qquad \qquad \qquad
\left . 
- 2f^{ba_1c}h_1^c h_+^{a_2} \ldots h_+^{a_{m+1}} 
\, \right]\\ &&\;\;+ \left.\,
\left[if^{ba_1c}j_0^c h_+^{a_2} \ldots h_+^{a_{m+1}}\right]
\, \, \right\}\,. 
\eeaa 
As before, there is some work to be done to show
that other terms, besides those written above, vanish along the way.
The second $[0]$ in the middle line is due to a total derivative.
Gathering together the terms containing the
factors $m$ yields, after a little rearranging,
\[
m\left\{ {\bf d}\left(j_+^2 h_+^{m-1}\right) - 2i\underline{h_+}
j_+ h_+^{m-1} \right\}\,.
\]
Now we add the rest of the terms, which are
\[
-i \underline{(3j_0 -2i h_1+b_1)}h_+^m -i 
m(b_0-b_1-2ih_0)\underline{h_+}h_+^{m-1}
=
-i {\bf d}\left((b_0-b_1-2ih_0)h_+^m\right) - 2i \underline{j_+} h_+^m
\,.
\]
The end result is 
\[
-2i\underline{j_+} h_+^m -  2im j_+\underline{h_+} h_+^{m-1} 
= -2i{\bf d}\left(j_+ h_+^m\right)=0 \,.
\]

\section{Appendix: Invariant tensors}

This section is a comprehensive guide to the invariant tensors of
relevance to this paper. As above, ${\cal G}$ is a Lie group with 
algebra ${\bf g}$
and $\rank({\bf g})=l$. Fundamental representation generators are $\{t_a\}$
and they satisfy 
\be
  \tr(t_a t_b) = -\de_{ab}\,, \quad [t_a,t_b] = f_{abc} t_c\,.
\ee
An arbitrary tensor $\Theta_{a_1 \dots a_n}$ is called invariant if we
have
\be
  \sum_{k=1}^n f^c_{\ ba_k} \Theta_{a_1 \dots a_{k-1} c a_{k+1} \dots
    a_n} = 0 \,. 
\ee
An equivalent statement is that the element of the enveloping algebra of
$\Lg$ given by
\be
  \hat{\Theta} = \Theta_{a_1 a_2 \dots a_n} t_{a_1} t_{a_2} \dots
    t_{a_n}\, , 
\ee
is a Casimir, that is $[t_b,\hat{\Theta}]=0$ for every $t_b$.
Symmetrizations and antisymmetrizations are denoted by
\be
\Theta_{(a_1 \ldots a_n )} = {1 \over n!} 
\sum_\sigma \Theta_{ a_{\sigma(1)} \ldots a_{\sigma(n)} }
\ , \qquad
\Theta_{[a_1 \ldots a_n ]} = {1 \over n!} 
\sum_\sigma (-1)^\sigma \Theta_{ a_{\sigma(1)} \ldots a_{\sigma(n)} }
\ee
respectively, where the sums extend over all permutations $\sigma$ of 
$\{ 1, 2, \dots , n \}$ and $(-1)^\sigma$ denotes the signature
of the permutation.

\subsection{Primitive symmetric tensors}

There are $l$ primitive symmetric tensors for each algebra ${\bf g}$. 
What this means is that any
symmetric invariant tensor can be expressed as a sum of tensor products
of these primitive tensors. The primitive tensors are not unique, but
the ambiguity in their selection consists of the freedom to add 
or subtract symmetrized tensor products, of the form 
\be
  d_{a_1 \dots a_n} = u_{(a_1 \dots a_k} v_{a_{k+1} \dots a_n)},
  \label{compound}
\ee
(up to overall constants).
In section 2 we introduced the term `compound' for such tensors.
For the classical algebras, we 
can take all but one of the primitive tensors to be 
\be\label{symtr}
  s_{a_1 a_2 \dots a_n} = 
\str \left( t_{a_1} t_{a_2} \dots t_{a_n} \right) = 
{\rm Tr} \left( t_{(a_1} t_{a_2} \dots t_{a_n)} \right)\, , 
\ee
where $n$ takes the values
\[
  \begin{array}{rl}
  a_l= su(l\!+\!1)\hspace{0.2in} & 2,3,\dots (l+1) \\ 
  b_l= so(2l\!+\!1)\hspace{0.2in} & 2,4, \dots 2l \\
  c_l= sp(2l)\hspace{0.2in} & 2,4, \dots 2l \\
  d_l= so(2l)\hspace{0.2in} & 2,4, \dots (2l-2) 
  \end{array}
\]
Note that this only defines $(l-1)$ tensors for the algebras $d_l$. The
final invariant in this case is the Pfaffian, given by
\be
  p_{a_1\dots a_l} = {1 \over 2^l \, l !} \, 
\eps^{j_1\dots j_{2l}} (t_{a_1})_{j_1j_2} \dots
    (t_{a_l})_{j_{2l-1}j_{2l}}\,. 
\ee

To illustrate these ideas, consider the algebra $a_3$. This algebra has
rank three and a set of primitive symmetric tensors is provided by
$$
  \tr(t_{a_1} t_{a_2})\,, \quad \str(t_{a_1} t_{a_2} t_{a_3})\,,
    \quad \str(t_{a_1} t_{a_2} t_{a_3} t_{a_4})\,. 
$$
The six-fold symmetric trace is non-primitive and can be written
\beaa
  \str (t_{a_1} t_{a_2} t_{a_3} t_{a_4} t_{a_5} t_{a_6}) &=&
    \frac{1}{3} \;\tr (t_{(a_1} t_{a_2} t_{a_3}) \;\tr
    (t_{a_4} t_{a_5} t_{a_6)}) \\ 
  &+& \frac{3}{4} \;\tr (t_{(a_1} t_{a_2}) \;\tr (t_{a_3} t_{a_4} t_{a_5}
    t_{a_6)}) \\
  &-& \frac{1}{8} \;\tr (t_{(a_1} t_{a_2}) \;\tr (t_{a_3} t_{a_4})
    \;\tr (t_{a_5} t_{a_6)}) \,.
\eeaa

\subsection{$\Omega$ tensors}

Given any symmetric invariant tensor $d^{(n)}_{a_1 \dots a_n}$ we can define
an order $(2n-1)$ antisymmetric invariant tensor by
\bea
  \Omega^{(2n-1)}_{a_1 a_2 \dots a_{2n-1}} &=& {1 \over 2^{n-1} } \,
    f^{b_1}_{\ [a_1a_2}
    f^{b_2}_{\ a_3a_4} \dots f^{b_{n-1}}_{\ a_{2n-3}a_{2n-2}} {d^{b_1
    b_2 \dots b_{n-1}}}_{a_{2n-1}]} \nonumber \\
  &=& {1 \over 2^{n-1}} \, f^{b_1}_{\ a_1[a_2} f^{b_2}_{\ a_3a_4} \dots
    f^{b_{n-1}}_{\ a_{2n-3}a_{2n-2}]} {d^{b_1 b_2 \dots
    b_{n-1}}}_{a_{2n-1}}.
  \label{cocycle}
\eea
The second equality follows from careful use of invariance
properties \cite{azca97}. We also observe that for any symmetric 
invariant tensor $d$,
\be
  d_{a_1 \dots a_n} f^{a_1}_{\ [b_1b_2} \dots f^{a_n}_{\
    b_{2n-1}b_{2n}]} = 0 , 
  \label{vanish}
\ee
by using invariance and the Jacobi identity.
This leads to a very useful property of the $\Omega$ tensors: if 
$d$ in (\ref{cocycle}) is compound,
as in (\ref{compound}), then $\Omega$ vanishes identically.
This is easily understood by considering 
the symmetrization of indices in (\ref{compound}) to be 
written out explicitly, followed by substitution in (\ref{cocycle}).
In every one of the resulting terms either 
$u$ or $v$ has all its indices contracted with structure constants, 
as in (\ref{vanish}), and the result follows. 
Since (\ref{cocycle}) is linear in $d$, the expression
for $\Omega$ will also vanish if $d$ is any sum of compound tensors. The
corollary to this is that only the primitive part of $d$ contributes to
$\Omega$.
From the primitive symmetric tensors (\ref{symtr}) we obtain
\be
\Omega_{a_1 \ldots a_{2n-1}} = \tr \left(t_{[a_1}
    t_{a_2} \ldots t_{a_{2n-1}]} \right) 
\ee
providing $l$ primitive antisymmetric tensors for $a_l$, $b_l$ and $c_l$.
For $d_l$, we have $l-1$ tensors of this form and the final
primitive antisymmetric tensor is that defined from the Pfaffian via
(\ref{cocycle}). In general, we have precisely $l$ primitive totally
antisymmetric invariant tensors, which are in 1-1  correspondence with
the primitive symmetric tensors of $\Lg$. 

\subsection{$\Lambda$ tensors}

Given a general symmetric invariant tensor $d^{(n)}_{a_1a_2 \dots a_n}$  we
define a $\Lambda$ tensor by
\be
  \Lambda^{(2n-2)}_{a_1 a_2 \dots a_{2n-2}} = 
{1 \over 2^{n-2}} \, f^{b_1}_{\ [a_1a_2} \dots
    f^{b_{n-2}}_{\ a_{2n-5}a_{2n-4}} {d^{b_1
    b_2 \dots b_{n-2}}}_{a_{2n-3}]a_{2n-2}}.
\ee
As these tensors have less symmetry than the $\Omega$ tensors we would
expect there to be a larger class of them. In the above section, we saw
that only the primitive part of the $d$-tensor contributed to
$\Omega$. We would like to know what is the analogue of this for the
$\Lambda$ tensors. The answer is that compound tensors 
(\ref{compound}) may contribute to $\Lambda$, but not if they 
can be written as a product of three of more factors:
\be
  d_{a_1 \ldots a_n} = u_{(a_1 \ldots a_r} v_{a_{r+1} \ldots a_s} 
w_{ a_{s+1} \ldots a_n)}
  \label{tricompound}
\ee
As before, we need only think of such a compound tensor used in
the definition of $\Lambda$ above, with the symmetrization on its 
indices written out. In each of the resulting terms, at least one of the 
constituents $u$, $v$ or $w$ will have all its indices contracted 
with structure constants, and so will vanish by (\ref{vanish}).

It remains to understand how compound tensors involving just 
two primitive constituents contribute to $\Lambda$.
This involves nothing more than substituting
a general tensor of this type 
\be
  d_{a_1 \dots a_p b_1 \ldots b_q} 
= d^{(p)}_{(a_1 \dots a_p} d^{(q)}_{b_{1} \dots b_q)}
\ee
into the definition. Some care is required with 
combinatorial factors, however, in order to arrive
at the result
\be\label{key}  
\Lambda^{(2n)}_{a_1 \ldots a_{2n-1} b}
= 
{pq \over (p{+}q)(p{+}q{-}1)} \, \left ( \, 
\Omega^{(2p-1)}_{ [ a_1 \ldots a_{2p-1}}
\Omega^{(2q-1)}_{a_{2p} \ldots a_{2n-1} ] b } 
+
\Omega^{(2q-1)}_{ [ a_1 \ldots a_{2q-1}}
\Omega^{(2p-1)}_{a_{2q} \ldots a_{2n-1} ] b }
\right ) 
\ee
where $n=p+q-1$ and $\Omega^{(2p-1)}$ and $\Omega^{(2q-1)}$ are related to
$d^{(p)}$ and $d^{(q)}$ as in (\ref{cocycle}).
Unlike the $\Omega$ tensors, there is not a unique
$\Lambda$ tensor for each primitive symmetric invariant. 
Nevertheless, we see that $\Lambda$ tensors based
on different $d$ tensors will differ only by linear 
combinations of products of $\Omega$ tensors, as in the 
expression above.

\subsection{Comments}

There are various awkward coefficients which arise in checking 
some statements made in section 4.2 relating  
to the bosonic terms in the definition (\ref{curlyK}) 
of the currents $\K^+_{m+1}$ for $su(N)$.
It was found that commuting charges 
could be obtained by modifying the 
current $\B^+_{m+1}$ by an expression including
\be -{m\over N} \sum_{p+q=m+1} \B^-_p \B^-_q
\ee
Considering first how these quantities can be written in terms of 
symmetric tensors, the modification amounts to changing 
$s^{(m+1)}_{a_1 a_2 a_3 \ldots a_{m+1}} j_+^{a_1} j_+^{a_2} h_+^{a_3}
\ldots h_+^{a_{m+1}}$ by 
\bea
&&-{1 \over N} \sum_{p+q = m+1}
(\, s^{(p)}_{a_1 a_2 \ldots a_p} \, j_+^{a_1} h_+^{a_2} \ldots h_+^{a_p}\, ) 
\, 
(\, s^{(q)}_{b_1 b_2 \ldots b_q} \, j_+^{b_1} h_+^{b_2} \ldots h_+^{b_q}\, ) 
\nonumber\\
&&= -{1 \over N} \sum_{p+q = m+1} 
{(p{+}q)(p{+}q{-}1) \over 2 pq} \, 
s^{(p)}_{(a_1 a_2 \ldots a_p} \,
s^{(q)}_{b_1 b_2 \ldots b_q)} \,
j_+^{a_1} j_+^{b_1} \, h_+^{a_2} \ldots h_+^{a_{p}} \,
h_+^{b_2} \ldots h_+^{b_{q}} \qquad \qquad 
\label{symod}
\eea
where care must be taken with symmetrizations in order 
to obtain the correct coefficents in the second expression.
This is easily found to 
reproduce the compound terms with two primitive factors 
which appear in the tensors $k^{(m+1)}$ 
listed in (\ref{suN}) in section 2.
Alternatively, in terms of antisymmetric tensors, we need to modify 
$\Lambda^{(2m)}_{a_1 a_2 \ldots a_{2m-1} b}\,  
j^{a_1}_+ \psi^{a_2}_+ \ldots \psi^{a_{2m-1}}_+ j^b_+ $
by the expression
\be\label{antisymod} 
-{1 \over N} \, {m \over 2m{-}1}
\sum_{p+q = m+1}
(\, \Omega^{(2p-1)}_{a_1 a_2 \ldots a_{2p-1} } \, 
j^{a_1}_+ \, \psi_+^{a_2} \ldots \psi_+^{a_{2p-1}} \, ) \,  
(\, \Omega^{(2q-1)}_{b_1 b_2 \ldots b_{2q-1} } \, 
j_+^{b_1} \, \psi_+^{b_2} \ldots \psi_+^{b_{2q-1}} \, ) \ . 
\ee
Now it can be checked that
\bea
&&
\left ( \, \Omega^{(2p-1)}_{[ a_1 a_2 \ldots a_{2p-1}}
\, \Omega^{(2q-1)}_{a_{2p} \ldots a_{2m-1} ] b } 
+
\Omega^{(2q-1)}_{ [a_1 a_2 \ldots a_{2q-1}}
\Omega^{(2p-1)}_{a_{2q} \ldots a_{2m-1} ] b } \, \right ) 
j^{a_1}_+ \, \psi_+^{a_2} \ldots \psi_+^{a_{2m-1}} j^b_+ 
\nonumber \\[3pt]
&=& {2p{+}2q{-2} \over 2p{+}2q{-}3} \, 
(\, \Omega^{(2p-1)}_{a_1 a_2 \ldots a_{2p-1} } \, 
j_+^{a_1} \, \psi_+^{a_2} \ldots \psi_+^{a_{2p-1}} \, ) \,  
(\, \Omega^{(2q-1)}_{b_1 b_2 \ldots b_{2q-1} } \, 
j_+^{b_1} \, \psi_+^{b_2} \ldots \psi_+^{b_{2q-1}} \, ) 
\nonumber
\eea
from which we see that (\ref{antisymod}) is equal to
\[
-{1 \over N} \, \sum_{p=2}^{m-1}
\Omega^{(2p-1)}_{[a_1 \ldots a_{2r-1} } \, 
\Omega^{(2m-2p+1)}_{a_{2p} \ldots a_{2m-1}] b } \, 
j^{a_1}_+ \, \psi_+^{a_2} \ldots \psi_+^{a_{2m-1}} \, j_+^b \ , 
\]
precisely as required for (\ref{mod}).
Notice also that this is consistent with (\ref{symod}) above on using
(\ref{key}).



\begin{thebibliography}{99}
\raggedright
\bibitem{EHMM2}
J.M. Evans, M. Hassan, N.J. MacKay, and A.J. Mountain,
{\em Local conserved charges in principal chiral models}, 
Nucl.~Phys.~{\bf B561} (1999) 385; {\tt hep-th/9902008}. 
%
\bibitem{EHMM1}
J.M. Evans, M. Hassan, N.J. MacKay, and A.J. Mountain,
{\em Conserved charges and supersymmetry in principal chiral models},
based on conference talks by JME, NJM and AJM; {\tt hep-th/9711140v3} 
(January 1999).
%
\bibitem{corri94}
E.~Corrigan,
\newblock{\em Recent developments in affine Toda quantum field theory},
\newblock Lectures given at CRM-CAP Summer School on Particles and
Fields '94, Banff, Canada, 16-24 Aug 1994,
\newblock preprint DTP-94/55; {\tt hep-th/9412213}.
%
\bibitem{dorey91} P.E. Dorey, {\em Root systems and purely elastic
S-matrices}, Nucl. Phys. {\bf B358} (1991) 654
%
\bibitem{chari95} 
V. Chari and A. Pressley, {\em Yangians, integrable
quantum systems and Dorey's rule}, Commun.~Math.~Phys.~{\bf 181} (1996)
265; {\tt hep-th/9505085}.
%
\bibitem{ogie86}
E. Ogievetsky, N. Reshetikhin and P. Wiegmann, {\em The principal
chiral field in two dimensions on classical Lie algebras: the Bethe
ansatz solution and factorized theory of scattering}, 
Nucl. Phys. {\bf B280} (1987) 45.
%
\bibitem{WZW} 
E. Witten, {\em Non-Abelian bosonization in two dimensions\/},
Commun.~Math.~Phys.~{\bf 92} (1984) 455.
%
\bibitem{balog90}
J. Balog, L. Feh\'er, L. O'Raifeartaigh, P. Forg\'acs, A. Wipf,
{\em Toda theory and W-algebra from a gauged WZNW point of view},
Ann. Phys. {\bf 203} (1990) 76.
%
\bibitem{azca97}
J.A. de Azc\'arraga, A.J. Macfarlane, A.J. Mountain and 
J.C. P\'erez Bueno, {\em Invariant tensors for simple groups},
Nucl. Phys. {\bf B510} (1998) 657; {\tt physics/9706006}.
%
\bibitem{AJM98} A.J. Mountain, 
{\em Invariant tensors and Casimir operators for simple compact Lie groups\/},
J.~Math Phys.~{\bf 39} (1998) 5601; {\tt physics/9802012}.
%
\bibitem{mack92}
N.J.~MacKay, {\em On the classical origins of Yangian symmetry 
in integrable field theory}, Phys. Lett. {\bf B281} (1992), 90;
err. ibid. {\bf B308} (1993) 444.
%
\bibitem{brez79}
E. Br\'ezin, C. Itzykson, J. Zinn-Justin and J.-B. Zuber,
{\em Remarks on the existence of non-local charges in two-dimensional
models}, Phys. Lett. {\bf 82B} (1979) 442.
%
\bibitem{Bern}
D.~Bernard, {\em On symmetries of some massless 2D field theories},
\newblock Phys. Lett. {\bf B279} (1992) 78.
%
\bibitem{Ferr78} S.~Ferrara, L.~Giradello and S.~Sciuto,
{\em An infinite set of conservation laws in the supersymmetric 
sine-Gordon theory},
Phys.~Lett.~{\bf B76} (1978) 303.
%
\bibitem{HP92} P.S.~Howe and G.~Papadopoulos, 
{\em Holonomy groups and W-symmetries}, 
Commun.~Math.~Phys.~{\bf 151} (1993) 467; {\tt hep-th/9202036}.
%
\bibitem{A}
E.~Abdalla, M.C.B.~Abdalla, J.C.~Brunelli and A.~Zadra,
{\em The algebra of non-local charges in non-linear sigma models},
\newblock Commun. Math. Phys. {\bf 166} (1994) 166.

M.C.B.~Abdalla,
{\em Integrability of chiral nonlinear sigma models with a
Wess-Zumino term},
\newblock Phys. Lett. {\bf B152} (1985) 215.
%
\bibitem{deV}
H.~J. de~Vega,
{\em Field theories with an infinite number of conservation
laws and B\"acklund transformations in two dimensions},
\newblock Phys.~Lett.~{\bf B87} (1979) 233.
%
\bibitem{SPCM} 
A.V.~Mikhailov, {\em Integrability of supersymmetric 
generalization of classical chiral models in two-dimensional space-time},
JETP Lett.~{\bf 28} (1978) 512.

Z.~Popowicz and L.-L.~Chau Wang, {\em Backlund transformation, 
local and non-local conservation laws for super-chiral fields}, 
Phys.~Lett.~{\bf B98} (1981) 253.

E.~Abdalla and M.~Forger, {\em Integrable non-linear $\sigma$-models 
with fermions}, Commun.~Math.~Phys.~ {\bf 104} (1986) 123.
%
\bibitem{EH} J.M.~Evans and T.J.~Hollowood, {\em Exact scattering in the 
SU(N) supersymmetric principal chiral model}, Nucl.~Phys.~{\bf B493} 
(1997) 517; {\tt hep-th/9603190}.
%
\bibitem{DKPR} 
P. Di Vecchia, V.G. Knizhnik, J.L. Petersen and P. Rossi, {\em A
supersymmetric Wess-Zumino lagrangian in two dimensions},
Nucl. Phys. {\bf B253} (1985) 701.
%
\bibitem{SWZW} J.~Fuchs, 
{\em Superconformal Ward identities and the WZW model}, 
Nucl.~Phys.~{\bf B286} (1987) 455;
{\em More on the super WZW theory}, Nucl.~Phys.~{\bf B318} (1989) 631.
%
\bibitem{gold80}
Y.Y. Goldschmidt and E. Witten, {\em Conservation laws in
some two-dimensional models}, Phys. Lett. {\bf B91} (1980) 392.
%
\bibitem{clark81}
T. E. Clark, S. T. Love and S. Gottlieb,
{\em Infinite number of conservation laws in two-dimensional
superconformal models},
Nucl. Phys. {\bf B186} (1981) 347.
%
\bibitem{curt80}
T. Curtright and C. Zachos, {\em Nonlocal currents for supersymmetric
nonlinear models}, Phys. Rev. {\bf D21} (1980) 411;
{\em Supersymmetry and the nonlocal Yangian deformation symmetry},
Nucl. Phys. {\bf B402} (1993) 604.

E. Corrigan and C. Zachos, {\em Non-local charges for the
supersymmetric $\sigma$-model}, Phys. Lett. {\bf B88} (1979) 273.
%
\bibitem{chau86}
L.-L. Chau and H. C. Yen,
{\em Integrability of the super-chiral model 
with a Wess-Zumino term},
Phys. Lett. {\bf B177} (1986) 368.
%
\bibitem{Pap} G.~Papadopoulos, {\em Supersymmetric Toda field theories},
Phys.~Lett.~{\bf B365} (1996) 98; {\tt hep-th/9508175}.
%
\bibitem{EM} J.M.~Evans and J.O.~Madsen, {\em Integrability vs Supersymmetry},
Phys.~Lett.~{\bf B389} (1996) 665; {\tt hep-th/9608190}.
%

\end{thebibliography}
\end{document}